# Coherent effects in the stochastic electrodynamics of two-fluid plasma

S.K.H. Auluck[1],[2]

## Abstract

Random electromagnetic fields are ubiquitous in plasmas, the most common example being electromagnetic radiation of thermal origin. They should exert a random force on electrons and ions in a plasma, adding a random component to their motion. Products of randomly fluctuating quantities, such as velocity and magnetic field, which are correlated through the dynamical equations of the two-fluid model of plasma, should then exhibit non-zero average values. Investigation of such effects requires spatial-spectral representation of the non-linear equations of the two-fluid model. Chandrasekhar-Kendall (CK) functions, their generating function and its gradient defined over an infinite domain are shown to simultaneously provide orthogonal basis for solenoidal, scalar and irrotational fields respectively, facilitating transformation from coordinate space to mode number space and back. This paper constructs a theoretical framework for studying coherent effects of random forces due to random electromagnetic fields in a two-fluid plasma and discusses some results which follow from its structure. Azimuthally symmetric modes are shown to be the *sole* beneficiaries of the cooperation of random modes in generating non-random effects. As a brief demonstration, it is shown that the convective electron acceleration $(\vec{v}_e \cdot \vec{\nabla})\vec{v}_e$ and the Lorentz acceleration resulting from random electromagnetic field have a non-zero average value for an idealized "just born", finite, azimuthally symmetric plasma interacting with its thermal radiation. This injects canonical momentum at different rates for ions and electrons into azimuthally symmetric modes, leading to spontaneous generation of both magnetic field and vorticity. The formalism also facilitates investigation of the interaction between compressible dynamics, which plays a central role in plasma compression, heating and confinement and incompressible dynamics, which is involved in phenomena like turbulence and self-organization.

[1] Homi Bhabha National Institute. Mumbai 400094. E-mail: skhauluck@gmail.com; skauluck@barc.gov.in
[2] Physics Group, Bhabha Atomic Research Center, Mumbai 400085, India.

I. Introduction:

Random electromagnetic (EM) fields are ubiquitous in plasmas. The most common example of such random EM fields is the thermal radiation, which constitutes a significant fraction of total energy content and transport in many astrophysical and fusion plasmas. The effect of thermal radiation on the plasma is usually treated in the theory of radiation hydrodynamics [1], which takes into account transport of radiation by absorption, re-emission and streaming, modification of its spectral form by free-free, free-bound, bound-bound and cyclotron radiation processes, its various interactions with particles and waves in the plasma and the contribution of the energy and momentum of radiation to the dynamics of plasma. Turbulence produced in plasmas driven far from equilibrium is another example of random EM fields.

Random EM fields and plasma could also interact in a different manner than studied under radiation hydrodynamics conditions. Electrons and ions of the plasma could be subjected to a randomly fluctuating force because of the random electromagnetic field, which would be superimposed on the non-random force from the electromagnetic field generated by charges and currents arising during evolution of the plasma. This would add a random component to their motion. Products of randomly fluctuating fields dependent on this motion, which are correlated through the equations of motion of the electrons and ions, would then have non-zero average values, representing coherent effects of random phenomena. While radiation hydrodynamics works with the energy-and-momentum-rich high wave-number part of the thermal radiation spectrum, such coherent effects are expected from low wave-number end of spectra of random EM fields comparable with reciprocal of size of the plasma. One effect of this kind, leading to amplification of a small "seed" axial magnetic field and generation of rotation in azimuthally symmetric plasma was discussed earlier [2] . These effects are likely to be important in some cases and development of a theoretical framework for their calculation should be a fruitful endeavor.

A theory of this type would find useful parallels in the theory known as Stochastic Electrodynamics [3], which explores the consequences of the interaction of a classical random EM radiation, assumed to be homogeneous, isotropic and Lorentz invariant, with classical point-like charged particles in the framework of Newtonian mechanics via the Lorentz force. The stipulation of Lorentz invariance leads to a unique spectrum identical with that found in quantum mechanics as the zero-point radiation. It has been shown that various effects, usually attributed to quantum behavior of fields and particles, can be reproduced within this theory, provided the scale factor for this random radiation is chosen to be proportional to the Planck constant. The SED approach represents the random EM field as superposition of a discrete spectrum of spatial eigenmodes (plane waves) in a rectangular cavity, with complex random amplitude which satisfies certain auto-correlation properties, and adds this source-less random EM field to the EM field generated by charges and currents while calculating forces. Usually, a limit is taken as the dimensions of the cavity tend to infinity, so that the spectrum of eigenmodes becomes continuous.

Construction of an analogous theory applicable to interaction between plasma and associated random EM fields requires two pre-requisites: (1) expansion of the equations of the two-fluid plasma model in a continuous spectrum of spatial eigenmodes in the natural (usually cylindrical) geometry of the plasma and (2) representation of the random electromagnetic field in terms of the same eigenmodes.

Spectral expansion methods in cylindrical geometry have been used extensively in plasma physics [4-12], usually in the context of non-linear magnetohydrodynamics (MHD) in physically bounded geometries, representing toroidal z-pinches or spheromaks. The presence of a conducting boundary close to the plasma is a necessary feature of these devices since currents induced in the conducting boundary play a significant role in plasma dynamics. This has two consequences for the theory: (1) The mode spectrum is discrete because of boundary conditions (2) The method is limited to incompressible modes.

These consequences arise as follows: Chandrasekhar-Kendall (CK) functions [13] (eigenfunctions of curl in cylindrical geometry) provide a complete orthogonal basis [14] for flux-less solenoidal fields in simply-connected cylindrical domains with periodic axial boundaries provided their radial component vanishes on the cylindrical boundary. This condition determines the discrete radial mode spectrum. The radial component of gradient of the associated scalar potential solution of the Helmholtz equation, which is the generating function of CK functions, is then non-zero on the boundary and cannot provide an orthogonal basis for irrotational vector fields, which must then be neglected as part of a physical simplifying assumption.

There are important plasma configurations, such as astrophysical plasmas, some varieties of linear z-pinches [15] and dense plasma focus [16], where a close-in conducting boundary is not present and the above referred limitation is unnecessary. On the other hand, non-linear interaction of compressible and incompressible electron modes can lead to new effects [2], such as spontaneous generation of axial magnetic field and rotation in azimuthally symmetric plasma, which are normally neglected but may represent a clue to understanding the high fusion reaction rate in plasma focus and z-pinches [17].

This paper therefore proposes a spatial-spectral representation with a continuous mode spectrum, in which CK functions, their generating function and its gradient defined over *infinite domain* provide orthogonal basis for solenoidal, scalar and irrotational fields respectively. A stochastic model of random electromagnetic field is constructed in terms of this basis using a model auto-correlation function.

The importance of including electron inertia has been highlighted earlier [2]. A subset of two-fluid equations with electron inertia, namely equations of continuity and momentum conservation, is considered. The energy conservation equations are omitted from the discussion for the time being. This is because they require a more detailed treatment of a different kind, which would distract from the issues related to equations of continuity and conservation of momentum.

The energy conservation equations have several aspects. Firstly, they determine how the pressure gradient force in the momentum conservation equations is related to changes in density brought about by compressible flow. Processes which locally add energy to the plasma particles, such as resistive heating and work done in compression and which take energy out of the plasma particles, such as radiation, heat conduction and collisions between species, determine how much energy remains available to increase temperature of the particle species. The temperature and density then determine the pressure through the Equation of State (EoS).

Secondly, they account for the partitioning of energy deposited by the external energy source between kinetic and thermal energy of particles and EM fields.

The inclusion of random fluctuations adds one more category to this function: partitioning of energy between random and non-random components of fields. This issue has important ramifications, mentioned later, which can be meaningfully discussed only after the structure of stochastic electrodynamics based on equations of continuity and conservation of momentum is explored in detail.

As a temporary measure, the detailed treatment of energy conservation is deferred and a polytropic EoS with index 2 is assumed for both ions and electrons. This converts the pressure gradient force from being an independent phenomenon to a function of density alone – a drastic simplification meant to reduce mathematical clutter in the present discussion *deliberately limited* to equations of continuity and momentum conservation.

Two-fluid equations of continuity and momentum conservation along with Maxwell's equations are then transformed into time-evolution equations in mode-number space. An ansatz for separation of random and non-random components of spectral coefficients of physical fields is proposed. Coherent effects of random EM fields are then isolated as additional terms in the spectral time-evolution equations.

This paper has a specific focus and limited scope related to coherent effects of random EM fields in two-fluid plasmas. The rich variety of plasma phenomena contained in the two-fluid model continues to have a latent presence in the formalism developed in the paper, although their explicit recognition and discussion would be out of place in the present context.

This paper aims to establish the mathematical framework of spatial-spectral representation of two-fluid plasma dynamics and random EM field in infinite geometry of basis functions, its adaptation to the case of random EM field interacting with the plasma and general observations on the structure of the source terms. Usually, compressible dynamics, which plays a central role in plasma compression, heating and confinement, and incompressible dynamics, which is involved in phenomena like turbulence and self-organization, are treated as stand-alone phenomena, using techniques which work with only one of them. The formalism being considered in this paper hopes to treat both compressible and incompressible dynamics on an equal footing so that their synergistic link can be explored meaningfully.

It is shown that spectrally-represented two-fluid equations of continuity and momentum conservation acquire additional source terms, which are quadratic in the amplitude of random EM field. Quite general physical assumptions lead to the conclusion that a finite two-fluid plasma interacting with its associated random EM field would always spontaneously create magnetic field and vorticity. A formalism for calculation of plasma evolution in the presence of random EM fields, which explicitly includes various kinds of dynamo effects, is the major outcome of this endeavor. Further numerical exploration in problem-specific terms is beyond the scope of this paper. Some elements of this mathematical framework have been established elsewhere, which are briefly recapitulated.

A brief overview of this paper is given below. The orthogonal basis functions for scalar, solenoidal and irrotational fields are defined in section II, which also presents their orthogonality properties in the form of pairs of integral transform relations enabling transformations from coordinate space to mode number space (also referred as transform space) and back.

These transform relations are used in Section III to translate the equations of continuity and momentum conservation for ions and electrons as well as Maxwell's equations into transform space. Presence of quadratic nonlinearity in these equations leads to nineteen functions of mode numbers involving infinite coordinate-space integration over product of three basis-functions along with various combinations of differential and vector operators. These functions contain all the physics represented by the differential and vector operators of the two-fluid equations. They also make frequent use of infinite integral over product of three Bessel functions of first kind of integer order, which is shown to have well-characterized singular behavior and for which, a semi-empirical Dirac-delta-function model has been proposed [18]. Certain physical assumptions, which reflect the finite nature of plasma and take into account the limits of validity of the two-fluid plasma model, are introduced.

Section IV describes a stochastic model for random electromagnetic field in terms of various correlations, with Planck radiation serving as one concrete example.

This electromagnetic field jiggles electrons and ions and introduces a random component in the electron and ion densities and velocities, which are correlated through the system of equations. An ansatz is proposed in Section V for separation of random and non-random components of all spectrally represented fields using autocorrelation properties of the stochastic model. It is shown that the evolution equations for non-random components of spectral coefficients acquire source terms related to the random EM field. Azimuthally symmetric modes are shown to be the *sole* beneficiaries of the cooperation of random modes in generating non-random effects.

Section V presents a brief demonstration of this formalism for an idealized problem consisting of a "just-born" finite plasma interacting with its own thermal radiation. It is shown that canonical momentum is injected into azimuthally symmetric modes of the plasma by its interaction with thermal radiation at different rates for ions and electrons. This can spontaneously generate both magnetic field and ion vorticity, when both are initially absent.

Section VI discusses the observations and summarizes the conclusions.

II. Orthogonal basis for spectral expansion

*Dimensionless* eigenfunctions of curl in cylindrical geometry[3] (Chandrasekhar-Kendall functions) are defined over infinite domain as $\vec{\nabla} \times \vec{\chi} =سk\vec{\chi}, s = \pm 1$, $0 \leq k \leq \infty$,

---
[3] A simply-connected domain is assumed.

$$\vec{\chi}_{m\kappa\gamma s} = k^{-2}\exp(im\theta - i\kappa z)\begin{pmatrix} i\dfrac{1}{2}\gamma\hat{r}\left((sk+\kappa)J_{m+1}(\gamma r) + (sk-\kappa)J_{m-1}(\gamma r)\right) \\ +\dfrac{1}{2}\gamma\hat{\theta}\left((sk+\kappa)J_{m+1}(\gamma r) - (sk-\kappa)J_{m-1}(\gamma r)\right) \\ +\hat{z}\gamma^2 J_m(\gamma r) \end{pmatrix} \quad 1$$

The variable s labels right and left handed helical modes and $\gamma \equiv \sqrt{k^2 - \kappa^2}$. The orthogonality relation over *infinite domain* is

$$\int d^3\vec{R}\; \vec{\chi}^*_{m\kappa\gamma s} \cdot \vec{\chi}_{m'\kappa'\gamma's'} = 8\pi^2 k^{-2}\gamma \delta_{mm'}\delta_{ss'}\delta(\kappa'-\kappa)\delta(\gamma'-\gamma) \quad 2$$

The generating function for the Chandrasekhar-Kendall functions

$$\psi_{m k \gamma}(r,\theta,z) = J_m(\gamma r)\exp(im\theta - i\kappa z) \quad 3$$

satisfies the condition

$$\int d^3\vec{R}\, \psi^*_{m k \gamma}\psi_{m' \kappa'\gamma'} = 4\pi^2\gamma^{-1}\delta_{mm'}\delta(\kappa'-\kappa)\delta(\gamma'-\gamma) \quad 4$$

Its gradient satisfies the conditions

$$\int d^3\vec{R}\,\vec{\nabla}\psi^*_{m k \gamma} \cdot \vec{\nabla}\psi_{m'\kappa'\gamma'} = 4\pi^2 k^2\gamma^{-1}\delta_{m,m'}\delta(\kappa-\kappa')\delta(\gamma-\gamma') \quad 5$$

$$\int d^3\vec{R}\,\vec{\nabla}\psi^*_{m k \gamma} \cdot \vec{\chi}_{m'\kappa'\gamma's'} = 4\pi^2 i k^{-2}\delta_{m,m'}\{\kappa\gamma' - \gamma\kappa'\}\delta(\kappa'-\kappa)\delta(\gamma'-\gamma) = 0$$

$$\int d^3\vec{R}\,\vec{\nabla}\psi_{m k \gamma} \cdot \vec{\chi}_{m'\kappa'\gamma's'} = -i4\pi^2(-1)^m \delta_{m,-m'}k^{-2}\{\gamma\kappa' + \kappa\gamma'\}\delta(\kappa'+\kappa)\delta(\gamma-\gamma') = 0 \quad 6$$

*The orthogonality between gradient of the generating function and the C-K functions is an exact consequence of their structure and does not depend on their asymptotic behavior.*

All fields can then be expressed in a spatial-spectral expansion using time-dependent mode coefficients with $\psi_{m k \gamma}$ as basis for scalar fields $S(\vec{R},t)$, $k^{-1}\vec{\nabla}\psi_{m k \gamma}$ as basis for irrotational fields $\vec{F}_{IR}(\vec{R},t)$ and $\vec{\chi}_{m k \gamma s}$ as basis for solenoidal fields $\vec{F}_{SOL}(\vec{R},t)$, *all basis functions being dimensionless*:

$$S(r,\theta,z,t) = \sum_{m'=-\infty}^{\infty} \int_{-\infty}^{\infty} d\kappa' \int_{0}^{\infty} d\gamma' S_{m'}(\kappa',\gamma',t) \psi_{m'\kappa'\gamma'}(r,\theta,z)$$

$$S_m(\kappa,\gamma,t) = \frac{\gamma}{4\pi^2} \int_{0}^{2\pi} d\theta \int_{-\infty}^{\infty} dz \int_{0}^{\infty} rdr \psi^*_{m\kappa\gamma}(r,\theta,z) S(r,\theta,z,t)$$

$$\vec{F}_{IR}(r,\theta,z,t) = \sum_{m'=-\infty}^{\infty} \int_{-\infty}^{\infty} d\kappa' \int_{0}^{\infty} d\gamma' \mathcal{F}^{IR}_{m'}(\kappa',\gamma',t) k'^{-1} \vec{\nabla} \psi_{m'\kappa'\gamma'}(r,\theta,z)$$

$$\mathcal{F}^{IR}_m(\kappa,\gamma,t) = \frac{\gamma}{4\pi^2 k} \int_{0}^{2\pi} d\theta \int_{-\infty}^{\infty} dz \int_{0}^{\infty} rdr \vec{\nabla} \psi^*_{m\kappa\gamma}(r,\theta,z) \cdot \vec{F}_{IR}(r,\theta,z,t)$$

$$\vec{F}_{SOL}(r,\theta,z,t) = \sum_{s'} \sum_{m'=-\infty}^{\infty} \int_{-\infty}^{\infty} d\kappa' \int_{0}^{\infty} d\gamma' \mathcal{F}^{SOL}_{m's'}(\kappa',\gamma',t) \vec{\chi}_{m'\kappa'\gamma's'}(r,\theta,z)$$

$$\mathcal{F}^{SOL}_{ms}(\kappa,\gamma,t) = \frac{k^2}{8\pi^2 \gamma} \int_{0}^{2\pi} d\theta \int_{-\infty}^{\infty} dz \int_{0}^{\infty} rdr \vec{\chi}^*_{m\kappa\gamma s}(r,\theta,z) \cdot \vec{F}_{SOL}(r,\theta,z,t)$$

The pairs of integral transform relations 7-9 are *mathematical identities*, valid regardless of the details of the physical problem. They offer the enormous benefit of enabling replacement of vector differential operators with functions of ordinary numbers. They also contain the mathematical idealizations "zero" and "infinity", which need to be suitably interpreted [19] for construction of a physical theory.

Clearly, phenomena, which take place far away from the physical extent of a finite plasma, represented by a "zero" radial or axial mode number, or at scale lengths smaller than the Debye length, represented by "infinite" radial, axial or azimuthal mode number, cannot be considered in a theory of finite-sized plasma based on the two-fluid plasma model. The upper and lower limits of the integral transform relations 7-9 may need to be re-interpreted as *physically significant* large and small numbers, according to the requirements of the context. The resulting theory would then represent a *mathematically approximate* model, which can be compared with experimentally observed reality "within experimental errors". For example, the "zero" radial mode number may be of the order of $\gamma_0 \sim a^{-1}$ where a is the radial "size" of the plasma and "infinite" mode number may have to be taken as reciprocal of the Debye length: $\gamma_\infty \sim \lambda_D^{-1}$ when model equations exhibit divergence at mathematical zero or infinity. Similarly, the maximum value of azimuthal mode number may be $m_\infty \sim (\gamma \lambda_D)^{-1}$. Such divergences, commonly found in approximate physical models, indicate limitations of the approximation. As Morse and Feshbach [19] point out, *the very definitions* of the classical fields like density or electric field are simplified mathematical idealizations of a much more complex physical reality.

Relations 7-9 are used in the next section to translate two-fluid partial differential equations into transform space labeled by $\{m, s, \kappa, \gamma\}$. The reality of fields implies that the complex conjugate of a spectral coefficient satisfies the relation

$$\boldsymbol{Q}^*_{ms}(\kappa, \gamma, t) = (-1)^m \boldsymbol{Q}_{-ms}(-\kappa, \gamma, t). \qquad 10$$

III. Spectral expansion of fields

The two-fluid plasma model is described by the following equations [20][4]: Equation of continuity for ions and electrons, labeled by subscript $\iota$ (i or e) is:

$$\frac{\partial n_\iota}{\partial t} + \left(\vec{\nabla} \cdot n_\iota \vec{v}_\iota\right) = 0 \; ; \qquad 11$$

Momentum conservation for ions and electrons, (using scalar Braginskii friction between them) is described by

$$\frac{\partial \vec{v}_\iota}{\partial t} + \left(\vec{v}_\iota \cdot \vec{\nabla}\right)\vec{v}_\iota = \frac{e}{m_\iota} q_\iota \left(\vec{E} + \vec{v}_\iota \times \vec{B} - \eta \vec{J}\right) - \frac{\vec{\nabla} p_\iota}{m_\iota n_\iota}; \quad q_i = +1, q_e = -1; \qquad 12$$

The Maxwell equations relate the fields to charges and currents

$$\vec{\nabla} \cdot \vec{E} = e\varepsilon_0^{-1}\left(n_i - n_e\right)$$
$$\vec{\nabla} \times \vec{B} = \mu_0 \vec{J} + c^{-2}\frac{\partial \vec{E}}{\partial t}; \vec{\nabla} \times \vec{E} = -\frac{\partial \vec{B}}{\partial t} \qquad 13$$
$$\vec{J} = e\left(n_i \vec{v}_i - n_e \vec{v}_e\right)$$

---
[4] Full two-fluid energy conservation equations are omitted from the discussion and replaced with polytropic EoS.

These equations, along with problem-specific initial and boundary conditions, describe the coupled electron and ion dynamics.

The following short hand notation is used in the subsequent discussion: $\int' \Leftrightarrow \int_{-\infty}^{\infty} d\kappa' \int_{0}^{\infty} d\gamma'$, $\sigma \Leftrightarrow (\kappa, \gamma)$. The dynamical fields are represented in a spectral expansion, according to the templates 7-9, with the mode coefficients $\boldsymbol{n}_{m\iota}(\sigma, t)$ for number density, $\boldsymbol{w}_{ms\iota}(\sigma, t)$ for solenoidal component of velocity, $\boldsymbol{u}_{m\iota}(\sigma, t)$ for irrotational component of velocity, $\iota$ being the species label (i or e). The solenoidal and irrotational components of the term[5] $\vec{\nabla} p_\iota / m_\iota n_\iota$ are represented by the mode coefficients $\boldsymbol{p}^{S}_{ms\iota}(\sigma, t)$ and $\boldsymbol{p}^{I}_{m\iota}(\sigma, t)$ respectively. The electric field $\vec{E}$ is expressed as

$$\vec{E} = \sum_{s'} \sum_{m'=-\infty}^{\infty} \int' \boldsymbol{E}_{m's'}(\sigma', t) \vec{\chi}_{m'\sigma's'} - \sum_{m'=-\infty}^{\infty} \int' \boldsymbol{f}_{m'}(\sigma', t) \vec{\nabla} \psi_{m'\sigma'} \qquad 14$$

The last term in 14 is gradient of the electrostatic potential $\phi$, whose spectral coefficient is $\boldsymbol{f}_m(\sigma, t)$. Representation for magnetic field becomes

$$\vec{B} = \sum_{s'} \sum_{m'=-\infty}^{\infty} \int' \boldsymbol{B}_{m's'}(\sigma', t) \vec{\chi}_{m'\sigma's'} \qquad 15$$

Maxwell's equations yield the following relations

$$\boldsymbol{f}_m(\sigma, t) = e k^{-2} \varepsilon_0^{-1} \left( \boldsymbol{n}_{mi}(\sigma, t) - \boldsymbol{n}_{me}(\sigma, t) \right) \qquad 16$$

$$\frac{\partial \boldsymbol{B}_{ms}(\sigma, t)}{\partial t} = -sk \boldsymbol{E}_{ms}(\sigma, t) \qquad 17$$

$$\frac{\partial \boldsymbol{E}_{ms}(\sigma, t)}{\partial t} = sc^2 k \boldsymbol{B}_{ms}(\sigma, t) - \varepsilon_0^{-1} \boldsymbol{J}^{S}_{ms}(\sigma, t) \qquad 18$$

---

[5] In a general case, baroclinic source terms may arise because of different dynamics governing density and temperature (e.g. anisotropic thermal conduction). In this paper, use of polytropic EoS precludes this case.

$$\frac{\partial \boldsymbol{f}_m(\sigma,t)}{\partial t} = (\varepsilon_0 k)^{-1} \boldsymbol{J}_m^I(\sigma,t) \qquad 19$$

The quantities $\boldsymbol{J}_{ms}^S(\sigma,t), \boldsymbol{J}_m^I(\sigma,t)$ are mode coefficients of solenoidal and irrotational components of current density.

$$\boldsymbol{J}_{ms}^S(\sigma,t) = \frac{ek^2}{8\pi^2\gamma}\left(\begin{array}{l}\sum\limits_{s'}\sum\limits_{m'=-\infty}^{\infty}\sum\limits_{m''=-\infty}^{\infty}\int'\!\!\int''\!\left(\boldsymbol{w}_{m's'i}(\sigma',t)\boldsymbol{n}_{m''i}(\sigma'',t)-\boldsymbol{w}_{m's'e}(\sigma',t)\boldsymbol{n}_{m''e}(\sigma'',t)\right)\mathbb{P}^{(1)mm'm''ss'}_{\sigma\sigma'\sigma''}\\ +\sum\limits_{m'=-\infty}^{\infty}\sum\limits_{m''=-\infty}^{\infty}\int'\!\!\int''\!\left(\boldsymbol{u}_{m'i}(\sigma',t)\boldsymbol{n}_{m''i}(\sigma'',t)-\boldsymbol{u}_{m'e}(\sigma',t)\boldsymbol{n}_{m''e}(\sigma'',t)\right)k'^{-1}\mathbb{P}^{(2)mm'm''s}_{\sigma\sigma'\sigma''}\end{array}\right) \quad 20$$

$$\boldsymbol{J}_m^I(\sigma,t) = \frac{e\gamma}{4\pi^2 k}\left(\begin{array}{l}\sum\limits_{m'=-\infty}^{\infty}\sum\limits_{m''=-\infty}^{\infty}\int'\!\!\int''\!\left(\boldsymbol{u}_{m'i}(\sigma',t)\boldsymbol{n}_{m''i}(\sigma'',t)-\boldsymbol{u}_{me}(\sigma',t)\boldsymbol{n}_{m''e}(\sigma'',t)\right)k'^{-1}\mathbb{P}^{(3)mm'm''}_{\sigma\sigma'\sigma''}\\ +\sum\limits_{s'}\sum\limits_{m'=-\infty}^{\infty}\sum\limits_{m''=-\infty}^{\infty}\int'\!\!\int''\!\left(\boldsymbol{w}_{m's'i}(\sigma',t)\boldsymbol{n}_{m''i}(\sigma'',t)-\boldsymbol{w}_{m's'e}(\sigma',t)\boldsymbol{n}_{m''e}(\sigma'',t)\right)\mathbb{P}^{(4)mm'm''s'}_{\sigma\sigma'\sigma''}\end{array}\right) \quad 21$$

The equations of continuity become

$$\frac{\partial \boldsymbol{n}_{m\iota}(\sigma,t)}{\partial t} = -\frac{\gamma}{4\pi^2}\sum_{s'}\sum_{m'=-\infty}^{\infty}\sum_{m''=-\infty}^{\infty}\int'\!\!\int''\boldsymbol{w}_{m's'\iota}(\sigma',t)\boldsymbol{n}_{m''\iota}(\sigma'',t)\mathbb{P}^{(5)mm'm''s'}_{\sigma\sigma'\sigma''}$$
$$-\frac{\gamma}{4\pi^2}\sum_{m'=-\infty}^{\infty}\sum_{m''=-\infty}^{\infty}\int'\!\!\int''\boldsymbol{u}_{m'\iota}(\sigma',t)\boldsymbol{n}_{m''\iota}(\sigma'',t)\left[k'^{-1}\mathbb{P}^{(6)mm'm''}_{\sigma\sigma'\sigma''}-k'\mathbb{P}^{(7)mm'm''}_{\sigma\sigma'\sigma''}\right] \qquad 22$$

The equations of momentum conservation become

$$\frac{\partial \boldsymbol{w}_{ms\iota}(\sigma,t)}{\partial t} - \frac{q_\iota e}{m_\iota}\boldsymbol{E}_{ms}(\sigma,t) + \frac{\eta q_\iota e}{m_\iota}\boldsymbol{J}^S_{ms}(\sigma,t) + \boldsymbol{p}^S_{ms\iota}(\sigma,t) =$$

$$-\frac{k^2}{8\pi^2\gamma}\sum_{s',s''}\sum_{m'=-\infty}^{\infty}\sum_{m''=-\infty}^{\infty}\int'\int''\boldsymbol{w}_{m's'\iota}(\sigma',t)\boldsymbol{w}_{m''s''\iota}(\sigma'',t)\mathbb{P}^{(8)\,mm'm''ss's''}_{\sigma\sigma'\sigma''}$$

$$-\frac{k^2}{8\pi^2\gamma}\sum_{s'}\sum_{m'=-\infty}^{\infty}\sum_{m''=-\infty}^{\infty}\int'\int''k''^{-1}\boldsymbol{w}_{m's'\iota}(\sigma',t)\boldsymbol{u}_{m''\iota}(\sigma'',t)\mathbb{P}^{(9)\,mm'm''ss'}_{\sigma\sigma'\sigma''}$$

$$-\frac{k^2}{8\pi^2\gamma}\sum_{s''}\sum_{m'=-\infty}^{\infty}\sum_{m''=-\infty}^{\infty}\int'\int''k'^{-1}\boldsymbol{u}_{m'\iota}(\sigma',t)\boldsymbol{w}_{m''s''\iota}(\sigma'',t)\mathbb{P}^{(10)\,mm'm''ss''}_{\sigma\sigma'\sigma''}$$

$$-\frac{k^2}{8\pi^2\gamma}\sum_{m'=-\infty}^{\infty}\sum_{m''=-\infty}^{\infty}\int'\int''k''^{-1}k'^{-1}\boldsymbol{u}_{m'\iota}(\sigma',t)\boldsymbol{u}_{m''\iota}(\sigma'',t)\mathbb{P}^{(11)\,mm'm''s}_{\sigma\sigma'\sigma''}$$

$$+\frac{q_\iota e}{m_\iota}\frac{k^2}{8\pi^2\gamma}\sum_{s',s''}\sum_{m'=-\infty}^{\infty}\sum_{m''=-\infty}^{\infty}\int'\int''\boldsymbol{w}_{m's'\iota}(\sigma',t)\boldsymbol{B}_{m''s''}(\sigma'',t)\mathbb{P}^{(12)\,mm'm''ss's''}_{\sigma\sigma'\sigma''}$$

$$+\frac{q_\iota e}{m_\iota}\frac{k^2}{8\pi^2\gamma}\sum_{s''}\sum_{m'=-\infty}^{\infty}\sum_{m''=-\infty}^{\infty}\int'\int''\boldsymbol{u}_{m'\iota}(\sigma',t)k'^{-1}\boldsymbol{B}_{m''s''}(\sigma'',t)\mathbb{P}^{(13)\,mm'm''ss''}_{\sigma\sigma'\sigma''}$$

$$\frac{\partial \boldsymbol{u}_{m\iota}(\sigma,t)}{\partial t} + \frac{kq_\iota e}{m_\iota}\boldsymbol{f}_m(\sigma,t) + \frac{\eta q_\iota e}{m_\iota}\boldsymbol{J}_m^I(\sigma,t) + \boldsymbol{p}_{m\iota}^I(\sigma,t) =$$

$$-\frac{\gamma}{4\pi^2 k}\sum_{s',s''}\sum_{m'=-\infty}^{\infty}\sum_{m''=-\infty}^{\infty}\int'\int'' \boldsymbol{w}_{m's'\iota}(\sigma',t)\boldsymbol{w}_{m''s''\iota}(\sigma'',t)\mathbb{P}^{(14)mm'm''s's''}_{\sigma\sigma'\sigma''}$$

$$-\frac{\gamma}{4\pi^2 k}\sum_{s'}\sum_{m'=-\infty}^{\infty}\sum_{m''=-\infty}^{\infty}\int'\int'' k''^{-1}\boldsymbol{w}_{m's'\iota}(\sigma',t)\boldsymbol{u}_{m''\iota}(\sigma'',t)\mathbb{P}^{(15)mm'm''s'}_{\sigma\sigma'\sigma''}$$

$$-\frac{\gamma}{4\pi^2 k}\sum_{s''}\sum_{m'=-\infty}^{\infty}\sum_{m''=-\infty}^{\infty}\int'\int'' k'^{-1}\boldsymbol{u}_{m'\iota}(\sigma',t)\boldsymbol{w}_{m''s''\iota}(\sigma'',t)\mathbb{P}^{(16)mm'm''s''}_{\sigma\sigma'\sigma''}$$

$$-\frac{\gamma}{4\pi^2 k}\sum_{m'=-\infty}^{\infty}\sum_{m''=-\infty}^{\infty}\int'\int'' k''^{-1}k'^{-1}\boldsymbol{u}_{m'\iota}(\sigma',t)\boldsymbol{u}_{m''\iota}(\sigma'',t)\mathbb{P}^{(17)mm'm''}_{\sigma\sigma'\sigma''} \quad 24$$

$$+\frac{eq_\iota}{m_\iota}\frac{\gamma}{4\pi^2 k}\sum_{s',s''}\sum_{m'=-\infty}^{\infty}\sum_{m''=-\infty}^{\infty}\int'\int'' \boldsymbol{w}_{m's'\iota}(\sigma',t)\boldsymbol{B}_{m''s''}(\sigma'',t)\mathbb{P}^{(18)mm'm''s's''}_{\sigma\sigma'\sigma''}$$

$$+\frac{eq_\iota}{m_\iota}\frac{\gamma}{4\pi^2 k}\sum_{s''}\sum_{m'=-\infty}^{\infty}\sum_{m''=-\infty}^{\infty}\int'\int'' k'^{-1}\boldsymbol{u}_{m'\iota}(\sigma',t)\boldsymbol{B}_{m''s''}(\sigma'',t)\mathbb{P}^{(19)mm'm''s''}_{\sigma\sigma'\sigma''}$$

The pressure gradient term in 24 is modeled using the polytropic EoS with index 2 as $\boldsymbol{p}_{m\iota}^I(\sigma,t) = 2k\boldsymbol{n}_{m\iota}(\sigma,t)p_{0\iota}/m_\iota n_{0\iota}^2$. The symbols $\mathbb{P}^{(1)mm'm''ss's''}_{\sigma\sigma'\sigma''}\ldots\mathbb{P}^{(19)mm'm''s''}_{\sigma\sigma'\sigma''}$, ('projections of non-linear operators on orthogonal bases'), are defined below

$$\mathbb{P}^{(1)mm'm''ss'}_{\sigma\sigma'\sigma''} \equiv \int d^3\vec{R}\psi_{m''\kappa''\gamma''}\vec{\chi}^*_{m\kappa\gamma s}\cdot\vec{\chi}_{m'\kappa'\gamma's'}; \qquad \mathbb{P}^{(2)mm'm''s}_{\sigma\sigma'\sigma''} \equiv \int d^3\vec{R}\psi_{m''\kappa''\gamma''}\vec{\chi}^*_{m\kappa\gamma s}\cdot\vec{\nabla}\psi_{m'\kappa'\gamma'}$$

$$\mathbb{P}^{(3)mm'm''}_{\sigma\sigma'\sigma''} \equiv \int d^3\vec{R}\psi_{m''\kappa''\gamma''}\vec{\nabla}\psi^*_{m\kappa\gamma}\cdot\vec{\nabla}\psi_{m'\kappa'\gamma'}; \qquad \mathbb{P}^{(4)mm'm''s'}_{\sigma\sigma'\sigma''} \equiv \int d^3\vec{R}\psi_{m''\kappa''\gamma''}\vec{\nabla}\psi^*_{m\kappa\gamma}\cdot\vec{\chi}_{m'\kappa'\gamma's'}$$

$$\mathbb{P}^{(5)mm'm''s'}_{\sigma\sigma'\sigma''} \equiv \int d^3\vec{R}\psi^*_{m\kappa\gamma}\left(\vec{\chi}_{m'\kappa'\gamma's'}\cdot\vec{\nabla}\right)\psi_{m''\kappa''\gamma''}; \qquad \mathbb{P}^{(6)mm'm''}_{\sigma\sigma'\sigma''} \equiv \int d^3\vec{R}\psi^*_{m\kappa\gamma}\left(\vec{\nabla}\psi_{m'\kappa'\gamma'}\cdot\vec{\nabla}\right)\psi_{m''\kappa''\gamma''}$$

$$\mathbb{P}^{(7)mm'm''}_{\sigma\sigma'\sigma''} \equiv \int d^3\vec{R}\psi^*_{m\kappa\gamma}\psi_{m''\kappa''\gamma''}\psi_{m'\kappa'\gamma'}; \qquad \mathbb{P}^{(8)mm'm''ss's''}_{\sigma\sigma'\sigma''} \equiv \int d^3\vec{R}\vec{\chi}^*_{m\kappa\gamma s}\cdot\left(\vec{\chi}_{m'\kappa'\gamma's'}\cdot\vec{\nabla}\right)\vec{\chi}_{m''\kappa''\gamma''s''}$$

$$\mathbb{P}^{(9)mm'm''ss'}_{\sigma\sigma'\sigma''} \equiv \int d^3\vec{R}\vec{\chi}^*_{m\kappa\gamma s}\cdot\left(\vec{\chi}_{m'\kappa'\gamma's'}\cdot\vec{\nabla}\right)\vec{\nabla}\psi_{m''\kappa''\gamma''} \qquad \mathbb{P}^{(10)mm'm''ss''}_{\sigma\sigma'\sigma''} \equiv \int d^3\vec{R}\vec{\chi}^*_{m\kappa\gamma s}\cdot\left(\vec{\nabla}\psi_{m'\kappa'\gamma'}\cdot\vec{\nabla}\right)\vec{\chi}_{m''\kappa''\gamma''s''}$$

$$\mathbb{P}^{(11)mm'm''s}_{\sigma\sigma'\sigma''} \equiv \int d^3\vec{R}\vec{\chi}^*_{m\kappa\gamma s}\cdot\left(\vec{\nabla}\psi_{m'\kappa'\gamma'}\cdot\vec{\nabla}\right)\vec{\nabla}\psi_{m''\kappa''\gamma''}; \mathbb{P}^{(12)mm'm''ss's''}_{\sigma\sigma'\sigma''} \equiv \int d^3\vec{R}\vec{\chi}^*_{m\kappa\gamma s}\cdot\vec{\chi}_{m'\kappa'\gamma's'}\times\vec{\chi}_{m''\kappa''\gamma''s''} \qquad 25$$

$$\mathbb{P}^{(13)mm'm''ss''}_{\sigma\sigma'\sigma''} \equiv \int d^3\vec{R}\vec{\chi}^*_{m\kappa\gamma s}\cdot\vec{\nabla}\psi_{m'\kappa'\gamma'}\times\vec{\chi}_{m''\kappa''\gamma''s''} \qquad \mathbb{P}^{(14)mm'm''s's''}_{\sigma\sigma'\sigma''} \equiv \int d^3\vec{R}\vec{\nabla}\psi^*_{m\kappa\gamma}\cdot\left(\vec{\chi}_{m'\kappa'\gamma's'}\cdot\vec{\nabla}\right)\vec{\chi}_{m''\kappa''\gamma''s''}$$

$$\mathbb{P}^{(15)mm'm''s'}_{\sigma\sigma'\sigma''} \equiv \int d^3\vec{R}\vec{\nabla}\psi^*_{m\kappa\gamma}\cdot\left(\vec{\chi}_{m'\kappa'\gamma's'}\cdot\vec{\nabla}\right)\vec{\nabla}\psi_{m''\kappa''\gamma''} \qquad \mathbb{P}^{(16)mm'm''s''}_{\sigma\sigma'\sigma''} \equiv \int d^3\vec{R}\vec{\nabla}\psi^*_{m\kappa\gamma}\cdot\left(\vec{\nabla}\psi_{m'\kappa'\gamma'}\cdot\vec{\nabla}\right)\vec{\chi}_{m''\kappa''\gamma''s''}$$

$$\mathbb{P}^{(17)mm'm''}_{\sigma\sigma'\sigma''} \equiv \int d^3\vec{R}\vec{\nabla}\psi^*_{m\kappa\gamma}\cdot\left(\vec{\nabla}\psi_{m'\kappa'\gamma'}\cdot\vec{\nabla}\right)\vec{\nabla}\psi_{m''\kappa''\gamma''} \qquad \mathbb{P}^{(18)mm'm''s's''}_{\sigma\sigma'\sigma''} \equiv \int d^3\vec{R}\vec{\nabla}\psi^*_{m\kappa\gamma}\cdot\left(\vec{\chi}_{m'\kappa'\gamma's'}\times\vec{\chi}_{m''\kappa''\gamma''s''}\right)$$

$$\mathbb{P}^{(19)mm'm''s''}_{\sigma\sigma'\sigma''} \equiv \int d^3\vec{R}\vec{\nabla}\psi^*_{m\kappa\gamma}\cdot\left(\vec{\nabla}\psi_{m'\kappa'\gamma'}\times\vec{\chi}_{m''\kappa''\gamma''s''}\right)$$

Equations 16 - 25, are an exact mathematical consequence of applying transform relations 7-9 to the equations 11-13 of two-fluid plasma model. Together with specification of the initial state of the plasma (from which initial mode coefficients can be calculated), they describe an initial value problem, physically equivalent to the initial and boundary value problem (with "boundaries" at infinity) associated with equations 11-13. Each of the equations 22-24 describes the rate of change of an 'evolving mode', labeled by unprimed mode numbers on the left hand side, in terms of a linear source term and an aggregate effect of *pairs* of 'interacting modes', labeled by primed mode numbers on the right hand side. The physics contained in the vector and differential operators of the two-fluid plasma model is transferred to the mathematical structure of the functions $\mathbb{P}^{(1)mm'm''ss'}_{\sigma\sigma'\sigma''}\cdots\mathbb{P}^{(19)mm'm''s''}_{\sigma\sigma'\sigma''}$, which are discussed in the Appendix I. All of them have the common factor $\delta_{m,(m'+m'')}\delta(\kappa'+\kappa''-\kappa)$.

These functions repeatedly use the integral

$$\mathfrak{D}^{\gamma\,\gamma'\,\gamma''}_{mm'm''} \equiv \int_0^\infty r dr J_m(\gamma r) J_{m'}(\gamma' r) J_{m''}(\gamma'' r) \qquad 26$$

Some properties of this integral, whose analytical theory is not known, have been discussed elsewhere [18]. It is trivially invariant under permutations of the pairs $(m,\gamma),(m',\gamma'),(m'',\gamma'')$. It diverges when any one of the three radial mode number arguments $\gamma,\gamma',\gamma''$ equals the sum or difference of the other two. This implies that interaction of two waves with radial mode numbers $\gamma',\gamma''$ results in generation of two waves with radial mode numbers $\gamma'+\gamma''$ and $|\gamma'-\gamma''|$. Numerical experiments with a large but finite upper limit of integration have been used to infer and propose the following approximate semi-empirical "model" [18] for $\mathfrak{D}_{mm'm''}^{\gamma\gamma'\gamma''}$ in the sense of distribution with respect to $\gamma''$:

$$\mathfrak{D}_{mm'(m-m')}^{\gamma\gamma'\gamma''} \cong \frac{1}{2\sqrt{\gamma\gamma'}} \begin{cases} \delta(\gamma''-(\gamma-\gamma'))H(\gamma-\gamma') \\ +(-1)^{m-m'}\delta(\gamma''-(\gamma'-\gamma))H(\gamma'-\gamma) \\ +(-1)^{m'}\delta(\gamma''-(\gamma'+\gamma)) \end{cases} ; \quad \begin{aligned} H(x) &= 1 \quad x \geq 0 \\ &= 0 \quad x < 0 \end{aligned} \qquad 27$$

This model of $\mathfrak{D}_{mm'm''}^{\gamma\gamma'\gamma''}$ leads to simplification in the form of functions $\mathbb{P}^{(1)mm'm''ss'}_{\sigma\sigma'\sigma''} \cdots \mathbb{P}^{(19)mm'm''s''}_{\sigma\sigma'\sigma''}$, which is also included in Appendix I.

As discussed in the previous section, physical context must be separately inserted in the theoretical framework based on the integral transform relations 7-9. The configuration that is of interest here is one, where the plasma, although finite in extent, is *not limited by a physical boundary*: typical examples are astrophysical plasmas or z-pinches [15]. Such plasmas are clearly finite in size in the sense that an imaginary finite cylinder can always be found such that electromagnetic phenomena outside it do not affect the internal plasma dynamics. The physical requirement that phenomena outside this region be excluded from the theory is equivalent to the assertion that there exist lower bounds on radial and axial mode numbers corresponding to the "size"[6] of the plasma.

In this paper, this deference to physical context is implemented by inserting a positive model constant $\gamma_0$, related to radial size of the plasma, in the Dirac delta functions in 27:

---

[6] It is understood that for a plasma not constrained by a physical boundary, some ambiguity may exist concerning what precisely is the size of the plasma: this is a matter which can be dealt with by adopting some convention.

$$\mathfrak{D}_{mm'(m-m')}^{\gamma\,\gamma'\,\gamma''} \cong \frac{1}{2\sqrt{\gamma\gamma'}} \begin{Bmatrix} \delta(\gamma''-(\gamma-\gamma')-\gamma_0)H(\gamma-\gamma') \\ +(-1)^{m-m'}\delta(\gamma''-(\gamma'-\gamma)-\gamma_0)H(\gamma'-\gamma) \\ +(-1)^{m'}\delta(\gamma''-(\gamma'+\gamma)-\gamma_0) \end{Bmatrix} \qquad 28$$

This requires that the difference between radial mode numbers of interacting modes be at least $\gamma_0$ for them to make a contribution to plasma dynamics. Likewise, another model constant $\kappa_0$ related to the axial extent of the plasma is inserted in the common Dirac-delta function factor referred above, while using the expressions in Appendix I. Since the axial mode number takes positive as well as negative values, this factor is replaced by $\frac{1}{2}\delta_{m,(m'+m'')}(\delta(\kappa'+\kappa''-\kappa+\kappa_0)+\delta(\kappa'+\kappa''-\kappa-\kappa_0))$. The requirement that only scale lengths of physical quantities larger than the Debye length be taken into account is used to limit the azimuthal mode numbers to $|m| \leq m_\infty \sim \text{int}(\gamma\lambda_D)^{-1}$. These assumptions are similar in spirit to Galerkin approximation, where a spectral representation is truncated to a finite number of modes.

    The idea behind these model assumptions is to understand, in adequately simplified but meaningful manner, the consequences of finiteness of unbounded plasma. They are not unique assumptions. Other kinds of simplifying assumptions may certainly be possible: they are not considered in this paper. The ideal infinite case can always be recovered, if desired, by putting the model constants $\gamma_0, \kappa_0$ to zero. It is observed that putting $\gamma_0$ equal to zero leads to divergences, suggesting that there is some deeper physical meaning involved in this model constant. As in all models, these model assumptions, justified on physical grounds, are to be validated by comparison of model predictions with experiments.

## IV. Stochastic model of random electromagnetic field

The complex amplitudes[7] of random EM field, assumed to be isotropic, are expressed in terms of a function $e_{m's'}(\vec{L},\kappa',\gamma')$ of random variables $\vec{L}$ as

$$\vec{E}_R(r,\theta,z,t,\vec{L}) = \sum_{s'}\sum_{m'=-\infty}^{\infty}\int_{-\infty}^{\infty}d\kappa'\int_0^{\infty}d\gamma' e_{m's'}(\vec{L},\kappa',\gamma')\exp(ick't)\vec{\chi}_{m'\kappa'\gamma's'}(r,\theta,z)$$

$$\vec{B}_R(r,\theta,z,t,\vec{L}) = \sum_{s'}\sum_{m'=-\infty}^{\infty}\int_{-\infty}^{\infty}d\kappa'\int_0^{\infty}d\gamma' c^{-1}is' e_{m's'}(\vec{L},\kappa',\gamma')\exp(ick't)\vec{\chi}_{m'\kappa'\gamma's'}(r,\theta,z)$$

The Bessel function identity

$$\sum_{m=-\infty}^{\infty} J_m^2(x) = 1 \qquad\qquad 30$$

can be used to show that

$$\sum_{s}\sum_{m=-\infty}^{\infty} \vec{\chi}^*_{m\kappa\gamma s}(r,\theta,z)\cdot\vec{\chi}_{m\kappa\gamma s}(r,\theta,z) = 4\gamma^2 k^{-2} \qquad\qquad 31$$

Using angular brackets to denote ensemble average over the random variables, the energy density of random radiation is given by

$$\langle \mathcal{E} \rangle \equiv \frac{1}{2}\varepsilon_0 \langle \vec{E}_R^* \cdot \vec{E}_R \rangle + \frac{1}{2\mu_0} \langle \vec{B}_R^* \cdot \vec{B}_R \rangle$$

$$= \frac{1}{2}\varepsilon_0 \sum_{s,s'} \sum_{m,m'=-\infty}^{\infty} \int d\kappa' \int d\gamma' \int d\kappa \int d\gamma \, (s's+1) \qquad 32$$

$$\times \langle e_{ms}^*(\vec{L},\kappa,\gamma) e_{m's'}(\vec{L},\kappa',\gamma') \rangle \exp(-ic(k'-k)t) \vec{\chi}_{m\kappa\gamma s}^* \cdot \vec{\chi}_{m'\kappa'\gamma's'}$$

---

[7] It is shown later that time evolution of random components of spectral coefficients is governed by linear equations. Complex exponential in time is a more efficient way of treating these equations. When products of two random components are required for calculation of a non-random quantity, compliance with 10 is enforced.

The random mode amplitude in 29 is expressed as $e_{ms}(\vec{L},\kappa,\gamma) \equiv \mathbb{F}(\kappa,\gamma) \mathbb{R}_{ms}(\vec{L},\kappa,\gamma)$, where $\mathbb{R}_{ms}(\vec{L},\kappa,\gamma)$ is a *dimensionless* complex function of three independent random variables $\vec{L} = (L_x, L_y, L_z)$ in the interval $\{-\infty, \infty\}$ having dimensions of length, which is *postulated* to have the property:

$$\langle \mathbb{R}_{ms}^*(\vec{L},\kappa,\gamma) \mathbb{R}_{m's'}(\vec{L},\kappa',\gamma') \rangle = \delta_{mm'} \delta_{ss'} k^2 \delta(\kappa-\kappa') \delta(\gamma-\gamma') \qquad 33$$

The assumed isotropy of the random EM field demands that its amplitude be identical at any angle with respect to the z-axis of the cylindrical coordinate system indicating that $\mathbb{F}(\kappa,\gamma) \equiv \mathbb{F}(k)$. In the subsequent discussion, the quantity $\mathbb{F}(k)$ is referred as the stochastic amplitude.

Then it can be shown that

$$\langle \mathcal{E} \rangle = 2\pi\varepsilon_0 \int_0^\infty kdk \cdot k^2 (\mathbb{F}(k))^2 \qquad 34$$

$\mathbb{F}(k)$ may be determined by equating the expression for energy density as a function of mode numbers for a particular model of random EM fields with that given in 34. As an example, radiation in thermal equilibrium with an optically thick plasma has the Planck radiation spectrum. Equating energy density given by 34 with the energy density for the Planck radiation yields the relation

$$\mathbb{F}(k) = \mathcal{E}_0 \sqrt{\left\{1 + \frac{2}{\exp(\hbar ck/k_B T_r) - 1}\right\}} \qquad \mathcal{E}_0 \equiv \sqrt{\frac{\hbar c}{4\pi^3 \varepsilon_0}} = 5.3655 \times 10^{-9} \text{ Volt} - \text{m} \qquad 35$$

In 35, $T_r$ is radiation temperature, assumed to be uniform in space, and $k_B$ is Boltzmann constant.

The feasibility of such formulation of random EM field depends on the existence of a function $\mathbb{R}_{ms}(\vec{L}, \kappa, \gamma)$ which satisfies 33. This is realized by defining

$$\mathbb{R}_{ms}(\vec{L}, \kappa, \gamma) \equiv \frac{\sqrt{\gamma}}{2\pi} k \delta_{s,s_0} \exp(im\Theta - i\kappa L_z) J_m(\gamma L_r) (p(\vec{L}))^{-0.5} \qquad 36$$

In 36, $s_0$ is a dummy index, $p(\vec{L})$ is the probability distribution function of the random variables and

$$L_r = \sqrt{L_x^2 + L_y^2}, \Theta = \tan^{-1}(L_y/L_x) \qquad 37$$

This reproduces 33, independent of the probability distribution function. The above model 36 is also consistent with the following result obtained by applying 10 to 33

$$\left\langle \mathbb{R}_{ms}(\vec{L}, \kappa, \gamma) \mathbb{R}_{m's'}(\vec{L}, \kappa', \gamma') \right\rangle = (-1)^m \delta_{m,-m'} k^2 \delta_{s,s'} \delta(\kappa + \kappa') \delta(\gamma - \gamma') \qquad 38$$

The expectation values are seen to have the character of distributions, a consequence of the fact that physical fields are defined as integral transforms of the spectra.

Because of postulated isotropy of the random EM field, the three random variables $(L_x, L_y, L_z)$ should have identical distributions, which are assumed to be the normal distribution, with parameter $\Lambda$ representing variance:

$$p(\vec{L}) \equiv p_x(L_x) \cdot p_y(L_y) \cdot p_z(L_z) = (2\pi)^{-1.5} \Lambda^{-3} \exp(-L^2/2\Lambda^2)$$
$$= (2\pi)^{-1.5} \Lambda^{-3} \exp(-L_r^2/2\Lambda^2) \exp(-L_z^2/2\Lambda^2) \qquad 39$$

The random function $\mathbb{R}$ then has the following expectation value

$$\left\langle \mathbb{R}_{ms}\left(\vec{L},\kappa,\gamma\right)\right\rangle = \int_0^{2\pi} d\Theta \int_{-\infty}^{\infty} dL_z \int_0^{\infty} L_r dL_r\, p(\vec{L})\, \mathbb{R}_{ms}\left(\vec{L},\kappa,\gamma\right)$$

$$= \frac{\sqrt{\gamma}}{2\pi}(2\pi)^{-0.75} \Lambda^{-1.5} k \delta_{s,s_0} \int_0^{2\pi} d\Theta \exp(im\Theta) \int_{-\infty}^{\infty} dL_z \exp(-i\kappa L_z) \exp\left(-L_z^2/4\Lambda^2\right)$$

$$\times \int_0^{\infty} L_r dL_r \exp\left(-L_r^2/4\Lambda^2\right) J_m(\gamma L_r)$$

$$= 4(2\pi)^{-0.75} \sqrt{\pi}\sqrt{\gamma}\, \delta_{s,s_0}\delta_{m,0}\, k\Lambda^{1.5} \exp(-k^2\Lambda^2)$$

Since $\Lambda$ is a free parameter of the stochastic model of random EM field described above, its value can be arbitrarily chosen. Hence, although the expectation value of $\mathbb{R}_{ms}(\vec{L},\kappa,\gamma)$ is not identically zero, it can be made as small as desired by choosing a sufficiently large $\Lambda$.

The following results are discussed in Appendix II:

$$\left\langle \mathbb{R}_{ms}^{-1}(\vec{L},\kappa,\gamma)\mathbb{R}_{m's'}(\vec{L},\kappa',\gamma')\right\rangle = \delta_{m,m'}\delta_{s,s'}\delta(\kappa-\kappa')\delta(\gamma-\gamma')$$

$$\left\langle \mathbb{R}_{ms}^{-1}(\vec{L},\kappa,\gamma)\right\rangle = 0$$

$$\left\langle \mathbb{R}_{ms}^{-1}(\vec{L},\kappa,\gamma)\mathbb{R}_{m''s''}(\vec{L},\kappa'',\gamma'')\mathbb{R}_{m's'}(\vec{L},\kappa',\gamma')\right\rangle = 0$$

For the Planck radiation case, for $\hbar ck/k_B T_r \ll 1$, $\mathbb{F}(k) \approx \mathcal{E}_0\sqrt{2k_B T_r/\hbar ck}$, which diverges at small k. As discussed earlier, k needs to be limited to reciprocal of plasma size. Effects related to the Planck model of radiation are therefore expected to arise at the largest scale length in the system.

This treatment is general enough to be adapted to any model of random electromagnetic fields, assumed to be isotropic, such as turbulence, detailed opacity model of radiation hydrodynamics or Vlasov wave-particle interaction model, where an expression for the energy density is available in spectral form.

V. <u>Separation of random and non-random components in two-fluid spectral model</u>

The central idea of this paper is that *randomness in all physical fields has a common origin[8] because they are correlated through dynamical equations*. This requires that every spectral coefficient $\boldsymbol{q}_{ms}(\kappa,\gamma,t)$ be represented as the sum of a random component (denoted by an undertilde) and a non-random component (denoted by an underbar), *where the random component is proportional to the same dimensionless random function* $\mathbb{R}_{ms}(\vec{L},\kappa,\gamma)$:

$$\boldsymbol{q}_{ms}(\kappa,\gamma,t) \equiv \underset{\sim}{\boldsymbol{q}}_{ms}(\kappa,\gamma,t)\mathbb{R}_{ms}(\vec{L},\kappa,\gamma) + \underline{\boldsymbol{q}}_{ms}(\kappa,\gamma,t) \qquad 42$$

Note that complex conjugates of the random and non-random parts behave differently:

$$\underset{\sim}{\boldsymbol{q}}^{*}_{ms}(\kappa,\gamma,t) = \underset{\sim}{\boldsymbol{q}}_{-ms}(-\kappa,\gamma,t);\ \underline{\boldsymbol{q}}^{*}_{ms}(\kappa,\gamma,t) = (-1)^{m}\underline{\boldsymbol{q}}_{-ms}(-\kappa,\gamma,t) \qquad 43$$

The random component of the solenoidal electric field spectrum is thus the sum of random component of solenoidal electric field because of plasma dynamics and that related to the stochastic amplitude: $\underset{\sim}{\boldsymbol{E}}_{ms} + \mathbb{F}(k)\exp(ickt)$ and similarly for the magnetic field: $\underset{\sim}{\boldsymbol{B}}_{ms} + isc^{-1}\mathbb{F}(k)\exp(ickt)$. Each spectrally represented field and each of equations 22-24 can then be rewritten in terms of the ansatz 42 and decomposed into random and non-random components using the above stochastic model, by first taking its ensemble average and second by multiplying it with $\mathbb{R}^{-1}_{ms}(\vec{L},\kappa,\gamma)$ and then taking its ensemble average.

The main consequence of this procedure is the following conclusion: Correlated random fluctuations in the interacting modes provide a contribution *only to azimuthally symmetric part* of the rate of change of the non-random component of the evolving mode at the axial mode number $\kappa_0$, which has significant magnitude at radial mode number $\gamma_0$

---

[8] There could be random processes other than random EM fields such as binary collisions with neutrals, ionization-recombination processes which are neglected.

This is the joint effect of the factor $\frac{1}{2}\delta_{m,(m'+m'')}\left(\delta(\kappa'+\kappa''-\kappa+\kappa_0)+\delta(\kappa'+\kappa''-\kappa-\kappa_0)\right)$ which comes from spectral decomposition of the two-fluid equations and the model assumptions of Section III, relation 28 and the factor $(-1)^{m'}\delta_{m',-m'}k^2\delta_{s'',s}\delta(\kappa''+\kappa')\delta(\gamma''-\gamma')$ which comes from the stochastic model.

Correlated fluctuations are thus seen to evolve a part of the system towards the single-curl-eigenfunction state

$$k_0^{-2}\left\{\gamma_0\kappa_0 J_1(\gamma_0 r)\sin(\kappa_0 z)\hat{r}+\gamma_0 sk_0 J_1(\gamma_0 r)\hat{\theta}\cos(\kappa_0 z)+\gamma_0^2 J_0(\gamma_0 r)\cos(\kappa_0 z)\hat{z}\right\},$$

which is often encountered in theories of plasma relaxation and was used [17] in connection with a model for the spectral anisotropy of neutron emission in a plasma focus.

The second consequence is that *random fluctuations are affected only by the azimuthally symmetric part of the non-random component of spectrum at axial mode number* $\kappa_0$. This is a joint effect of the above referred common factor and relations 41.

A third consequence is that random modes at radial mode number $\gamma$ are affected by non-random modes at radial mode numbers $\gamma_0$ and $2\gamma-\gamma_0$. This comes from relations 28 and 41

The separated equation of continuity given below serves as a concise illustration of these statements and the structure of the resulting theoretical framework:

$$\frac{\partial \underline{\boldsymbol{n}}_{m\iota}(\kappa,\gamma,t)}{\partial t} = -\frac{\gamma}{4\pi^2}\sum_{s'}\sum_{m'=-\infty}^{\infty}\sum_{m''=-\infty}^{\infty}\int'\int''\mathbb{P}^{(5)\,mm'm''s'}_{\sigma\sigma'\sigma''}\underline{\boldsymbol{w}}_{m's'\iota}(\kappa',\gamma',t)\underline{\boldsymbol{n}}_{m''\iota}(\kappa'',\gamma'',t)$$

$$-\frac{\gamma}{4\pi^2}\sum_{m'=-\infty}^{\infty}\sum_{m''=-\infty}^{\infty}\int'\int''\left[k'^{-1}\mathbb{P}^{(6)\,mm'm''}_{\sigma\sigma'\sigma''} - k'\mathbb{P}^{(7)\,mm'm''}_{\sigma\sigma'\sigma''}\right]\underline{\boldsymbol{u}}_{m'\iota}(\kappa',\gamma',t)\underline{\boldsymbol{n}}_{m''\iota}(\kappa'',\gamma'',t)$$

$$-\frac{1}{8}\delta_{m,0}\{\delta(\kappa-\kappa_0)+\delta(\kappa+\kappa_0)\}\sqrt{\gamma}\sum_{m'=-(\lambda_D(\gamma+\gamma_0)/2)^{-1}}^{(\lambda_D(\gamma+\gamma_0)/2)^{-1}}(-1)^{m'}\int'\delta(\gamma'-(\gamma+\gamma_0)/2)$$

$$\times\underline{\boldsymbol{n}}^*_{m'\iota}(\kappa',\gamma',t)\left\{i\sqrt{\gamma'}(\kappa'\gamma-\kappa\gamma')\left(\sum_{s'}\underline{\boldsymbol{w}}_{m's'\iota}(\kappa',\gamma',t)\right) - \frac{\sqrt{2}k'(\kappa\kappa'+\gamma\gamma')\underline{\boldsymbol{u}}_{m'\iota}(\kappa',\gamma',t)}{\sqrt{(\gamma+\gamma_0)}}\right\}$$

$$-\frac{\sqrt{\gamma}}{4}\delta_{m,0}\{\delta(\kappa-\kappa_0)+\delta(\kappa+\kappa_0)\}\delta(\gamma-\gamma_0)\int'\sum_{m'=-(\gamma'\lambda_D)^{-1}}^{(\gamma'\lambda_D)^{-1}}$$

$$\times\underline{\boldsymbol{n}}^*_{m'\iota}(\kappa',\gamma',t)\left\{i\sqrt{\gamma'}(\kappa'\gamma-\kappa\gamma')\left(\sum_{s'}\underline{\boldsymbol{w}}_{m's'\iota}(\kappa',\gamma',t)\right) - k'\sqrt{\gamma'^{-1}}(\kappa\kappa'+\gamma\gamma')\underline{\boldsymbol{u}}_{m'\iota}(\kappa',\gamma',t)\right\}$$

$$\tag{44}$$

$$\frac{\partial \underline{\tilde{\boldsymbol{n}}}_{m\iota}(\kappa,\gamma,t)}{\partial t}$$

$$= \underline{\boldsymbol{n}}_{m\iota}(\kappa,\gamma,t)\frac{\sqrt{\gamma}}{4}\int'\{\delta(\kappa'+\kappa_0)+\delta(\kappa'-\kappa_0)\}\{\delta(\gamma'-\gamma_0)+(-1)^m\delta(\gamma'-(2\gamma-\gamma_0))\}$$

$$\times\left\{k'^{-1}\sqrt{\gamma'^{-1}}(\kappa\kappa'+\gamma\gamma')\underline{\boldsymbol{u}}_{0\iota}(\kappa',\gamma',t) - i\sqrt{\gamma'}k'^{-2}(\kappa'\gamma-\kappa\gamma')\sum_{s'}\underline{\boldsymbol{w}}_{0s'\iota}(\kappa',\gamma',t)\right\}$$

$$+\tfrac{1}{2}(i\kappa k^{-2}\gamma^2)(-1)^m\sum_{s}\underline{\tilde{\boldsymbol{w}}}_{ms\iota}(\kappa,\gamma,t)$$

$$\times\int''\delta(\gamma''-(2\gamma+\gamma_0))\{\delta(\kappa''+\kappa_0)+\delta(\kappa''-\kappa_0)\}\underline{\boldsymbol{n}}_{0\iota}(\kappa'',\gamma'',t)$$

$$+\frac{k^{-1}}{4}\underline{\tilde{\boldsymbol{u}}}_{m\iota}(\kappa,\gamma,t)\int''\underline{\boldsymbol{n}}_{0\iota}(\kappa'',\gamma'',t)\{\delta(\kappa''+\kappa_0)+\delta(\kappa''-\kappa_0)\}$$

$$\times\left\{2k^2\delta(\gamma''-\gamma_0)+(\kappa\kappa-\gamma\gamma)(-1)^m\delta(\gamma''-(2\gamma+\gamma_0))\right\}$$

$$\tag{45}$$

The following description of equations 44 and 45 applies equally well to other separated equations, which are given in Appendix III.

The evolution equation for each non-random spectral mode consists of two groups of terms. One group contains only non-random interacting modes: this represents the conventional equations in transform space, which describe the evolution of the plasma for a given initial configuration. The second group contains products of pairs of random spectral modes, one of which occurs as a complex conjugate. The second group contributes to evolution of the azimuthally symmetric part of non-random spectrum at the scale of the macroscopic size of the plasma and is proportional to the mean-square of the random EM field.

The evolution equations describing random parts of the spectra of electric and magnetic fields, density, solenoidal and irrotational velocity form a system of first order coupled linear equations (referred as the random system), with the stochastic electric field $\mathbb{F}(k)\exp(ickt)$ serving as inhomogeneous source term. The coefficients of the homogeneous terms depend only on non-random spectra at $m=0, \kappa=\pm\kappa_0, \gamma=\gamma_0$ and $2\gamma\pm\gamma_0$. It would be reasonable to assume that the non-random spectra evolve on a time scale which is much slower than the time scale of random spectra, so that they may be taken as adiabatic parameters in the random system. This then effectively becomes a mean-field theory for the evolution of the non-random spectra.

This system of first order coupled linear equations may show unstable growth under some conditions representing onset of turbulence. If so, the conditions leading to instability would depend only upon *azimuthally symmetric non-random spectra* at the largest scale lengths. Growing random spectra in turn would contribute to growth of non-random spectra for $m=0, \kappa=\kappa_0, \gamma=\gamma_0$: this is reminiscent of self-organization of a large scale structure arising from turbulence.

Solution of these equations for a particular plasma problem is beyond the scope of this paper. These equations can be transformed back into coordinate space using the transform relations 7-9, recovering the conventional equations with additional terms dependent on random radiation.

VI. <u>Spontaneous generation of canonical momentum in symmetric "just born" plasma</u>

As a brief demonstration of this formalism, the following idealized problem is considered. A finite plasma is created "instantaneously" along with its thermal radiation, assumed to have the Planck form. The random EM field then jiggles electrons, creating constant average *convective* electron acceleration $(\vec{v}_e \cdot \vec{\nabla})\vec{v}_e$ and Lorentz force $e\vec{v}_e \times \vec{B}_{rad}$, both of which depend on the mean-square of the random EM field.

The initial conditions are: all non-random spectral coefficients except that of density and pressure are zero. This also implies that there is no physical phenomenon which favors one value of the parameter s over the other, leading to degeneracy with respect to s. To the first order of Picard iteration [21], which describes evolution at early times, the random spectrum of solenoidal velocity of electrons is governed by the random EM field:

$$\underline{\boldsymbol{w}}_{mse}(\kappa,\gamma,t) = -\frac{e}{m_e}\frac{\mathbb{F}(k)\exp ickt}{(ick+\nu_0)} \qquad 46$$

Here $\nu_0$ is an effective collision frequency for the mode, which depends on the azimuthally symmetric spectral coefficient of density. It is assumed that $\nu_0 \ll \gamma_0 c$.

This is correlated with the random magnetic field $\underline{\boldsymbol{B}}_{ms}(\kappa,\gamma,t) = isc^{-1}\mathbb{F}(k)\exp(ickt)$. The non-random solenoidal velocity then obeys the following relation derived from A-28, A-29 and A-33:

$$\frac{\partial}{\partial t}\left(\underline{\boldsymbol{w}}_{ms\iota}(\kappa,\gamma,t) + \frac{q_\iota e}{m_\iota}\underline{\boldsymbol{A}}_{ms}(\kappa,\gamma,t)\right) + \frac{\eta q_\iota e}{m_\iota}\underline{\boldsymbol{J}}^S_{ms}(\kappa,\gamma,t)$$
$$= i\frac{23}{45}\frac{\mu_0 e^2 k_B T_r}{\pi^3 m_\iota^2 \lambda_D}\frac{k_0^2\kappa_0}{\gamma_0^2}\delta_{m,0}\delta(\gamma-\gamma_0)\{\delta(\kappa-\kappa_0) - \delta(\kappa+\kappa_0)\} \qquad 47$$

Since initially, $\underline{\boldsymbol{w}}_{ms\iota}, \underline{\boldsymbol{A}}_{ms}, \underline{\boldsymbol{J}}^S_{ms}$ are all zero, *the random radiation is seen to inject canonical momentum into the plasma at different rates for ions and electrons in azimuthally symmetric modes at early times*. As ions and electrons move in opposite directions, friction grows and leads to a quasi-steady state over the time scale $t_{qs} \sim \mu_0 \eta^{-1} k_0^{-2}$ for low magnetic Reynolds numbers typically encountered at early stages of plasma evolution:

$$\frac{\partial}{\partial t}\left(\underline{w}_{ms\iota}(\kappa,\gamma,t) + \frac{q_\iota e}{m_\iota}\underline{A}_{ms}(\kappa,\gamma,t)\right) \approx 0$$

$$\underline{J}_{ms}^{S}(\kappa,\gamma,t_{qs}) \approx -i\eta^{-1}\frac{23}{45}\frac{\mu_0 e k_B T_r}{\pi^3 m_e \lambda_D}\frac{k_0^2 \kappa_0}{\gamma_0^2}\delta_{m,0}\delta(\gamma-\gamma_0)\{\delta(\kappa-\kappa_0)-\delta(\kappa+\kappa_0)\}$$  48

Maxwell's equations give for this quasi-steady state

$$\underline{A}_{ms}(\kappa,\gamma,t_{qs}) = -\mu_0^2 i \frac{23}{45}\frac{k_B T_r}{\pi^3 \lambda_D}\frac{e}{\eta m_e}\delta_{m,0}\delta(\gamma-\gamma_0)\{\delta(\kappa-\kappa_0)-\delta(\kappa+\kappa_0)\}\left(\frac{\kappa_0}{\gamma_0^2}\right)$$  49

The non-random solenoidal velocity spectrum in the quasi-steady state is given by $\underline{w}_{ms\iota} = -q_\iota e \underline{A}_{ms}/m_\iota$ for both ions and electrons. The spontaneous magnetic field corresponding to 49 is

$$\vec{B}_{sp} = -2\frac{\mu_0^2 e}{\eta m_e}\frac{k_B T_r}{\pi^3 \lambda_D}\frac{\kappa_0 \sin(\kappa_0 z) J_1(\gamma_0 r)}{\gamma_0}\hat{\theta}$$  50

The magnetic field 50 has only the azimuthal component. A net axial component of magnetic field does not arise because of degeneracy with respect to s. A symmetry-breaking initial axial magnetic field can remove this degeneracy. It was shown [2] that a small seed axial magnetic field can grow because of non-linear interaction between compressible and incompressible electron modes driven by fluctuations. The magnetic moment of an electron stream [22] arising from its spin may provide the symmetry-breaking field which can be amplified by this mechanism [2].

This discussion of an idealized situation demonstrates *one* average effect of random EM field interacting with the plasma: simultaneous generation of magnetic field and fluid vorticity [23] *when both were initially absent*. Finiteness of the plasma plays a decisive role: for an axially infinite plasma, $\kappa_0 = 0$ and the effect vanishes. Infinite models of plasma, such as the Bennett z-pinch model, ignore such phenomena at the outset.

The spontaneous magnetic field and vorticity can act as seeds for dynamo mechanisms of various kinds powered by irrotational flow, all of which are present in the system of spectrally-represented two-fluid equations, leading to their amplification. For example, a convergent compressible flow can amplify both the solenoidal velocity and magnetic field as the magnetic Reynolds number grows at later stages of plasma evolution.

The above discussion suggests that *a finite two-fluid plasma and its thermally excited EM radiation will always be accompanied by magnetic field and vorticity*. This easily falsifiable prediction can form basis for experimental tests. Existing literature suggests some promising possibilities in this respect.

Stamper et.al. [24] have observed *azimuthal* magnetic field in plasma produced by laser ablation of a solid target. Briand et. al. [25] have observed *axial* magnetic field in a similar experiment. Stamper et.al. have interpreted their azimuthal magnetic field in terms of the baroclinic source term arising from axial temperature gradient and radial density gradient. Briand et. al. attribute their axial magnetic field to a turbulent dynamo effect. However, what determines the polarity of their axial magnetic field is not clear. Korobkin and Serov [26] have observed a "magnetic moment" in a laser produced spark which is perpendicular to the direction of laser beam. From the disposition of their probes, it is clear that they have observed azimuthal magnetic field. Similar experiments can be designed to look for the suggested coherent effects of random radiation and to discriminate from other possible mechanisms. The model will, however, have to be worked out in more detail, taking into account the possibility of amplification of the magnetic field and vorticity by macroscopic motion of the plasma.

VII. Summary and conclusions:

This paper attempts to construct a formalism which can calculate coherent effects of correlated random fluctuations of fields in finite, unbounded, two-fluid plasma, driven by associated random electromagnetic field, which may have any origin, magnitude or spectral form. This formalism is based on the following five ingredients in addition to the two-fluid plasma model: (1) An orthogonal basis for solenoidal, scalar and irrotational fields using Chandrasekhar-Kendall functions, their generating function and its gradient respectively, over infinite domain. (2) A semi-empirical model for the infinite integral over product of three Bessel functions of first kind of integer order. (3) A physical re-interpretation of mathematical "zero" and "infinity" in the context of a finite but unbounded plasma by placing limits on mode numbers. (4) A stochastic model for the random electromagnetic field described on the basis of CK functions in terms of a known formula for energy density in mode numbers space. (5) An ansatz for separation of random and non-random parts of field spectra.

These ingredients play the following role: (1) The orthogonal basis allows calculation of coupling between compressible and incompressible components of plasma dynamics: for example, fluctuations in charge density resulting from fluctuations in solenoidal electron velocity, which are directly driven by the random EM field. A non-zero average of product of these two quantities would act as a solenoidal current density, capable of creating magnetic field and inductive electric field, representing a new source term for EM fields. It also facilitates dealing with non-uniform density profile in a comprehensive manner.   (2) The semi-empirical formula for the triple-Bessel integral calculates resonant interaction between three radial waves. It shows generation of two radial waves at radial mode numbers equal to sum and difference of radial mode numbers of two interacting radial waves. The former represents generation of spectrum at smaller radial scale lengths as in shock steepening and turbulence; the latter represents generation of spectrum at a larger radial scale length as in self-organization. (3) The particular manner of introducing a physical interpretation of mathematical "zero" and "infinity" acts as a device to enable a simple but meaningful discussion of an extremely complicated process. Finiteness of the plasma is shown to play an important role: the calculated spontaneous magnetic field is zero for an axially infinite plasma model. The number of azimuthal modes needs to be limited to a finite value related to Debye length in order to avoid divergent results.  (4) The stochastic model of radiation provides a mathematical procedure for introducing a global random function in two-fluid plasma dynamics. (5) The ansatz for separation of random and non-random parts of spectra provides a mathematical tool for calculating random and non-random parts of any combination of physical fields.

When put together, these ingredients lead to azimuthally symmetric source terms in the evolution equations for *non-random* spectra in transform space with amplitude proportional to the radiation temperature or equivalently, average energy of random EM modes for a turbulent plasma. The *random* spectra form a system of coupled first order linear differential equations, with non-random spectra playing the role of adiabatic parameters. The two-fluid model then effectively acquires the character of a mean-field theory for non-random spectra: they are determined by a system of equations which contain terms which are of order zero, one and two in the non-random spectra. Such systems are capable of exhibiting phase-transition like behavior leading to spontaneous generation of one or more order parameters.

Early results from this formalism include the conclusion that coherent effects of correlated random fields drive a part of the plasma towards azimuthally symmetric curl-eigenfunction states at the largest scale-length. This suggests that the formalism could be a useful tool to look at relaxation phenomena, which also generate curl-eigenfunction states. A finite "just born" plasma is seen to spontaneously generate magnetic field and fluid vorticity. This example roughly corresponds to the transition zone between neutral gas and fully ionized plasma just ahead of the plasma sheath in a plasma focus, where the magnetic Reynolds number $R_m$ is small. Vorticity produced in this region would be amplified by the convergent radial flow of the sheath, (which has high $R_m$), once it reaches this zone. This could account for a significant fraction of plasma ions reaching a high directed kinetic energy while remaining confined within the plasma even when there are no instabilities [17].

This formalism treats both compressible dynamics, responsible for plasma compression, heating and confinement, and incompressible dynamics responsible for phenomena like turbulence and self-organization, on an equal footing. The random EM field, including that of thermal origin, is seen to provide a link between compressible dynamics and incompressible dynamics. It drives solenoidal random modes, whose solenoidal quadratic average effects can be amplified by convergent irrotational flow. The resulting dynamics would be very different from the dynamics predicted by conventional two-fluid model for given initial plasma configuration: containing *unexpected solenoidal flows and fields* seeded by the random EM field and amplified by the convergent irrotational flow. This points towards the following non-trivial conjecture: *it is impossible to have a purely compressible flow in a finite cylindrical two-fluid plasma*.

There is some experimental evidence in support of this conjecture.

The experimental diagnostics in the first case [27] comprised a *highly planar* diamagnetic loop consisting of a 2 mm wide, 20 micron thick and 100 mm diameter ring of copper machined out of a copper laminated printed circuit board. A gap of 0.5 mm was created in the ring, across which, a 50 Ohm coaxial cable was soldered with minimum lead length and terminated in 50 Ohms at the oscilloscope. The ring, along with its connection with the cable, was coated with a layer of epoxy resin to rule out any current due to photoemission of electrons or due to any other interaction with plasma particles. This ring could be mechanically mounted within a cylindrical vacuum chamber of a plasma focus perpendicular to and centered with respect to the axis with good accuracy. The plasma focus had a 22 mm diameter, 110 mm long anode, 86 mm outer diameter cathode consisting of 12 rods of 10 mm diameter and was operated at 70 kA peak current. The ring was placed at a height 10 mm above the anode and *was well outside the cathode. No signal is expected from such an experimental configuration.* Yet in a series of few tens of shots, a signal, representing rate of change of flux of axial magnetic field over the loop and clearly different in shape from the dI/dt signal, was observed in every shot [Fig. 1]. Moreover, the signal began with the start of current and continued long after the pinch phase, ruling out an m=1 MHD instability of a Bennett Z-pinch as a cause.

The second kind of plasma was a self-triggered low pressure spark [28], operating at a voltage in the range 1-5 kV, between two pointed electrodes kept 30-50 mm apart in a vacuum chamber filled with ~ 50-100 mbar of air or helium and directly mounted on a 2.8 µF low inductance energy storage capacitor. A highly planar diamagnetic loop of 80 mm diameter made with stiff copper wire was mounted perpendicular to and centered with respect to the axis. A signal very different from the dI/dt signal was observed in this loop [Fig. 2]. This signal was observed to continue for milliseconds although the dI/dt signal reached baseline after few microseconds.

These experimental observations, *which can be easily replicated in many laboratories*, are difficult to interpret in terms of current state of understanding and could possibly serve as test beds for the suggested coherent effects of random EM fields.

*This formalism is at present incomplete*: non-incorporation of energy conservation is only *one* instance of incompleteness. It is easily appreciated that quadratic quantities such as power density $\vec{J}\cdot\vec{E}$ and torque density $\vec{r}\times\vec{J}\times\vec{B}$ would also have a contribution from the random EM field. Flow of energy from the plasma particles to radiation is commonly known; a new dimension would be a reverse flow from radiation or turbulent random fields to bulk motion of plasma particles and to magnetic field over scale lengths comparable with plasma size *in an azimuthally symmetric manner*. Radiation may sustain a kind of current drive in some cases; phenomena such as ball lightening and the persistent of axial flux signal for milliseconds in the second experiment mentioned above may be examples of such current drive. Rate equilibrium between transfer of energy from particles to radiation and from radiation to particles may turn out to be an important aspect (in analogy with a similar phenomenon in SED): this needs a careful consideration. That is the reason why energy conservation is not addressed in the present paper. Determination of the model constants $\gamma_0$ and $\kappa_0$ within the theory in terms of some closure is also an important issue to be tackled.

Under special conditions, the system of linear equations for random fields may exhibit instability and the energy density of random EM fields may become much more than that of thermal radiation at low wave numbers. Investigation of non-random motion of ions and electrons arising from correlated random fluctuations in this case may represent a step towards understanding many strange phenomena [17] in finite unbounded plasmas like plasma focus and z-pinches such as the extremely high fusion reaction rate and asymmetry in space-resolved energy spectra of fusion products from laterally symmetric points, which have remained unexplained for lack of adequate theoretical efforts.

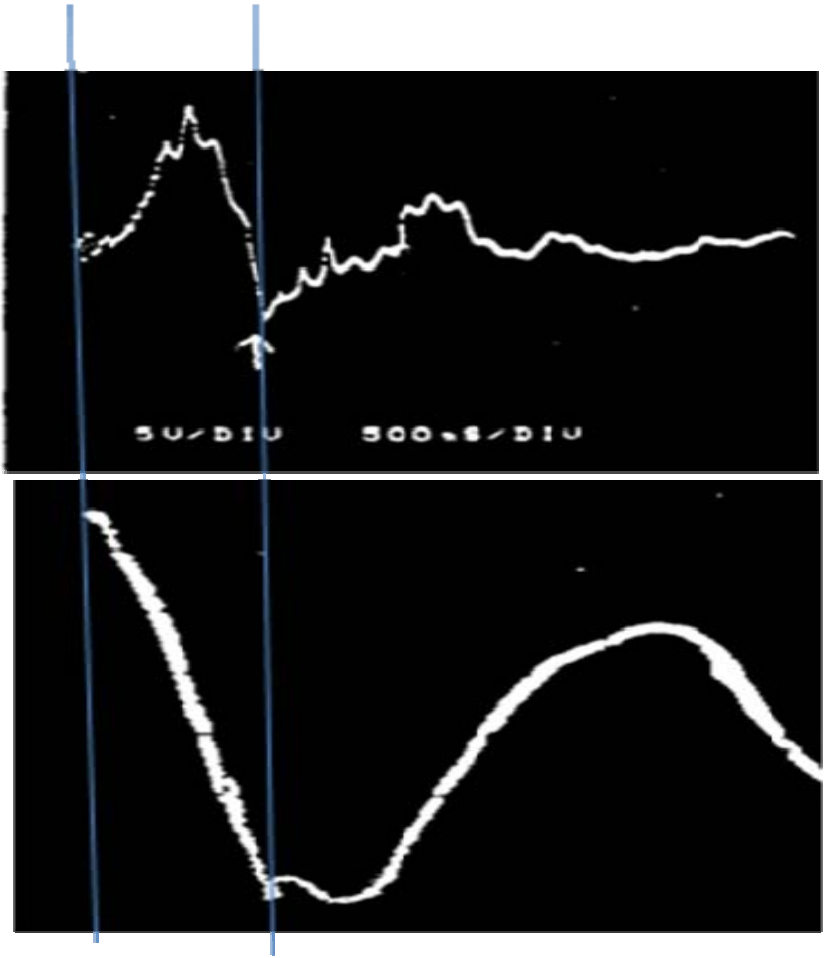

Fig. 1. Upper signal is from the diamagnetic loop. Lower signal is the dI/dt signal from a Rogowski coil. Note that the diamagnetic signal begins at the same time as the discharge current

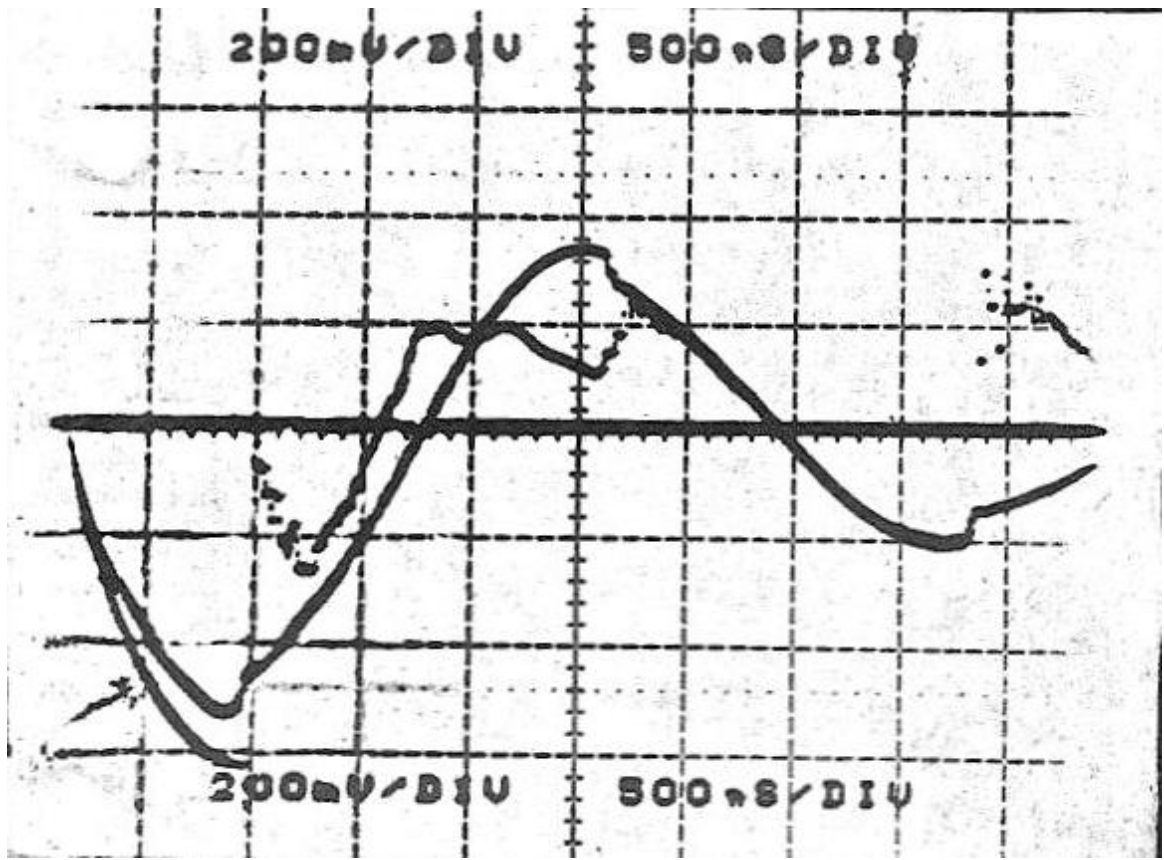

Fig 2. The signal marked with arrow is from the diamagnetic loop. The other signal is dI/dt signal from a magnetic probe.

## APPENDIX I

### Projection of non-linear operators on orthogonal bases

The integrals $\mathbb{P}^{(1)mm'm''ss's''}_{\sigma\sigma'\sigma''} \cdots \mathbb{P}^{(19)mm'm''s''}_{\sigma\sigma'\sigma''}$ defined in the main text can be evaluated in a straight forward manner in terms of the integral $\mathfrak{D}^{\gamma\gamma'\gamma''}_{mm'm''} \equiv \int_0^\infty rdr J_m(\gamma r) J_{m'}(\gamma' r) J_{m''}(\gamma'' r)$ to obtain the results given below. Also included is their simplified form using the approximate semi-empirical "model" for this integral

$$\mathfrak{D}^{\gamma\gamma'\gamma''}_{mm'(m-m')} \cong \frac{1}{2\sqrt{\gamma\gamma'}} \begin{Bmatrix} \delta(\gamma'' - (\gamma - \gamma'))H(\gamma - \gamma') \\ +(-1)^{m-m'}\delta(\gamma'' - (\gamma' - \gamma))H(\gamma' - \gamma) \\ +(-1)^{m'}\delta(\gamma'' - (\gamma' + \gamma)) \end{Bmatrix} ; \begin{matrix} H(x) = 1 & x \geq 0 \\ = 0 & x < 0 \end{matrix}$$

$$\mathbb{P}^{(1)mm'm''ss's''}_{\sigma\sigma'\sigma''} \equiv \int d^3\vec{R} \psi_{m''\kappa''\gamma''} \vec{\chi}^*_{m\kappa\gamma s} \cdot \vec{\chi}_{m'\kappa'\gamma's'}$$

$$= 4\pi^2 \delta_{m,(m'+m'')} \delta(\kappa' + \kappa'' - \kappa) k^{-2} k'^{-2}$$

$$\times \begin{Bmatrix} \frac{\gamma\gamma'}{2}(s'k' + \kappa')(sk + \kappa) \mathfrak{D}^{\gamma\ \gamma'\ \gamma''}_{(m+1)(m'+1)m''} \\ + \frac{\gamma\gamma'}{2}(s'k' - \kappa')(sk - \kappa) \mathfrak{D}^{\gamma\ \gamma'\ \gamma''}_{(m-1)(m'-1)m''} + \gamma^2\gamma'^2 \mathfrak{D}^{\gamma\gamma'\gamma''}_{mm'm''} \end{Bmatrix}$$

$$\cong 2\pi^2 \delta_{m,(m'+m'')} \delta(\kappa' + \kappa'' - \kappa) k^{-2} k'^{-2}$$

$$\times \sqrt{\gamma\gamma'} \begin{Bmatrix} \delta(\gamma'' - (\gamma - \gamma'))H(\gamma - \gamma')(ss'kk' + \kappa\kappa' + \gamma\gamma') \\ -(-1)^{m'}\delta(\gamma'' - (\gamma' + \gamma))(ss'kk' + \kappa\kappa' - \gamma\gamma') \\ +(-1)^{m-m'}\delta(\gamma'' - (\gamma' - \gamma))H(\gamma' - \gamma)(ss'kk' + \kappa\kappa' + \gamma\gamma') \end{Bmatrix}$$

A1

$$\mathbb{P}^{(2)\mathrm{mm'm''s}}_{\sigma\sigma'\sigma''} \equiv \int d^3\vec{R}\,\psi_{m''\kappa''\gamma''}\vec{\chi}^*_{m\kappa\gamma s} \cdot \vec{\nabla}\psi_{m'\kappa'\gamma'}$$

$$= 4\pi^2 i k^{-2}\delta_{m,(m'+m'')}\delta(\kappa'+\kappa''-\kappa)$$

$$\times\left\{\begin{array}{l}\dfrac{1}{2}\gamma\gamma'(sk+\kappa)\mathfrak{D}^{\gamma\quad\gamma'\quad\gamma''}_{(m+1)(m'+1)m''}\\[6pt] -\dfrac{1}{2}\gamma\gamma'(sk-\kappa)\mathfrak{D}^{\gamma\quad\gamma'\quad\gamma''}_{(m-1)(m'-1)m''}-\kappa'\gamma^2\mathfrak{D}^{\gamma\;\gamma'\;\gamma''}_{mm'm''}\end{array}\right\}$$

$$= 2\pi^2 i k^{-2}\delta_{m,(m'+m'')}\delta(\kappa'+\kappa''-\kappa)$$

$$\times\left\{\begin{array}{l}(\kappa\gamma'-\kappa'\gamma)\sqrt{\gamma\gamma'^{-1}}\,\delta(\gamma''-(\gamma-\gamma'))H(\gamma-\gamma')\\[4pt] -(\kappa\gamma'+\kappa'\gamma)\sqrt{\gamma\gamma'^{-1}}\,(-1)^{m'}\delta(\gamma''-(\gamma'+\gamma))\\[4pt] +(\kappa\gamma'-\kappa'\gamma)\sqrt{\gamma\gamma'^{-1}}\,(-1)^{m-m'}\delta(\gamma''-(\gamma'-\gamma))H(\gamma'-\gamma)\end{array}\right\}$$

A2

$$\mathbb{P}^{(3)\mathrm{mm'm''}}_{\sigma\sigma'\sigma''} \equiv \int d^3\vec{R}\,\psi_{m''\kappa''\gamma''}\vec{\nabla}\psi^*_{m\kappa\gamma}\cdot\vec{\nabla}\psi_{m'\kappa'\gamma'}$$

$$= 4\pi^2\delta_{m,(m'+m'')}\delta(\kappa'+\kappa''-\kappa)\left\{\begin{array}{l}\dfrac{\gamma\gamma'}{2}\mathfrak{D}^{\gamma\quad\gamma'\quad\gamma''}_{(m-1)(m'-1)m''}\\[6pt]+\dfrac{\gamma\gamma'}{2}\mathfrak{D}^{\gamma\quad\gamma'\quad\gamma''}_{(m+1)(m'+1)m''}+\kappa\kappa'\mathfrak{D}^{\gamma\;\gamma'\;\gamma''}_{mm'm''}\end{array}\right\}$$

$$= 2\pi^2\delta_{m,(m'+m'')}\dfrac{1}{\sqrt{\gamma\gamma'}}\delta(\kappa'+\kappa''-\kappa)$$

$$\times\left\{\begin{array}{l}(\gamma\gamma'+\kappa\kappa')\delta(\gamma''-(\gamma-\gamma'))H(\gamma-\gamma')\\[4pt]-(-1)^{m'}(\gamma\gamma'-\kappa\kappa')\delta(\gamma''-(\gamma'+\gamma))\\[4pt]+(-1)^{m-m'}(\gamma\gamma'+\kappa\kappa')\delta(\gamma''-(\gamma'-\gamma))H(\gamma'-\gamma)\end{array}\right\}$$

A3

$$\mathbb{P}^{(4)mm'm''s''}_{\sigma\sigma'\sigma''} \equiv \int d^3\vec{R}\,\psi_{m''\kappa''\gamma''}\vec{\nabla}\psi^*_{m\kappa\gamma}\cdot\vec{\chi}_{m'\kappa'\gamma's'}$$

$$= 4\pi^2 ik'^{-2}\delta_{m,(m'+m'')}\delta(\kappa'+\kappa''-\kappa)$$

$$\times\left\{\begin{array}{l}\dfrac{1}{2}\gamma\gamma'(s'k'-\kappa')\mathfrak{D}^{\gamma\ \gamma'\ \gamma''}_{(m-1)(m'-1)m''}\\[6pt] -\dfrac{1}{2}\gamma\gamma'(s'k'+\kappa')\mathfrak{D}^{\gamma\ \gamma'\ \gamma''}_{(m+1)(m'+1)m''}+\kappa\gamma'^2\mathfrak{D}^{\gamma\ \gamma'\ \gamma''}_{mm'm''}\end{array}\right\}$$

$$= 2\pi^2 ik'^{-2}\sqrt{\gamma'\gamma^{-1}}\,\delta_{m,(m'+m'')}\delta(\kappa'+\kappa''-\kappa)$$

$$\times\left\{\begin{array}{l}+(\kappa\gamma'-\kappa'\gamma)\delta(\gamma''-(\gamma-\gamma'))H(\gamma-\gamma')\\[4pt]+(\kappa'\gamma+\kappa\gamma')(-1)^{m'}\delta(\gamma''-(\gamma'+\gamma))\\[4pt]+(\kappa\gamma'-\kappa'\gamma)(-1)^{m-m'}\delta(\gamma''-(\gamma'-\gamma))H(\gamma'-\gamma)\end{array}\right\}$$

A4

$$\mathbb{P}^{(5)mm'm''s'}_{\sigma\sigma'\sigma''} \equiv \int d^3\vec{R}\,\psi^*_{m\kappa\gamma}\left(\vec{\chi}_{m'\kappa'\gamma's'}\cdot\vec{\nabla}\right)\psi_{m''\kappa''\gamma''}$$

$$= ik'^{-2}4\pi^2\delta_{m,(m'+m'')}\delta(\kappa'+\kappa''-\kappa)$$

$$\times\left\{\begin{array}{l}\dfrac{\gamma'\gamma''}{2}(s'k'+\kappa')\mathfrak{D}^{\gamma\ \gamma'\ \gamma''}_{m(m'+1)(m''-1)}\\[6pt]-(s'k'-\kappa')\dfrac{\gamma'\gamma''}{2}\mathfrak{D}^{\gamma\ \gamma'\ \gamma''}_{m(m'-1)(m''+1)}-\kappa''\gamma'^2\mathfrak{D}^{\gamma\ \gamma'\ \gamma''}_{mm'm''}\end{array}\right\}$$

$$= ik'^{-2}2\pi^2\sqrt{\gamma'\gamma^{-1}}\,\delta_{m,(m'+m'')}\delta(\kappa'+\kappa''-\kappa)$$

$$\times\left\{\begin{array}{l}(\kappa'\gamma-\kappa\gamma')\delta(\gamma''-(\gamma-\gamma'))H(\gamma-\gamma')\\[4pt]-(\kappa'\gamma+\kappa\gamma')(-1)^{m'}\delta(\gamma''-(\gamma'+\gamma))\\[4pt]+(\kappa'\gamma-\kappa\gamma')(-1)^{m-m'}\delta(\gamma''-(\gamma'-\gamma))H(\gamma'-\gamma)\end{array}\right\}$$

A5

$$\mathbb{P}^{(6)\mathrm{mm'm''}}_{\sigma\sigma'\sigma''} \equiv \int d^3\vec{R}\,\psi^*_{\mathrm{m\kappa\gamma}}\left(\vec{\nabla}\psi_{\mathrm{m'\kappa'\gamma'}}\cdot\vec{\nabla}\right)\psi_{\mathrm{m''\kappa''\gamma''}}$$

$$= -4\pi^2 \delta_{\mathrm{m,(m'+m'')}}\delta(\kappa'+\kappa''-\kappa)\left\{\begin{array}{l}\dfrac{\gamma'\gamma''}{2}\mathfrak{D}^{\gamma\ \gamma'\ \gamma''}_{\mathrm{m(m'-1)(m''+1)}}\\ +\dfrac{\gamma'\gamma''}{2}\mathfrak{D}^{\gamma\ \gamma'\ \gamma''}_{\mathrm{m(m'+1)(m''-1)}}+\kappa'\kappa''\mathfrak{D}^{\gamma\ \gamma'\ \gamma''}_{\mathrm{mm'm''}}\end{array}\right\}$$

$$= -\dfrac{2\pi^2}{\sqrt{\gamma\gamma'}}\delta_{\mathrm{m,(m'+m'')}}\delta(\kappa'+\kappa''-\kappa)$$

$$\left\{\begin{array}{l}(\kappa\kappa'+\gamma\gamma'-k'^2)\delta(\gamma''-(\gamma-\gamma'))\mathrm{H}(\gamma-\gamma')\\ +(\kappa\kappa'-\gamma'\gamma-k'^2)(-1)^{\mathrm{m'}}\delta(\gamma''-(\gamma'+\gamma))\\ +(\kappa\kappa'+\gamma\gamma'-k'^2)(-1)^{\mathrm{m-m'}}\delta(\gamma''-(\gamma'-\gamma))\mathrm{H}(\gamma'-\gamma)\end{array}\right\}$$

A6

$$\mathbb{P}^{(7)\mathrm{mm'm''}}_{\sigma\sigma'\sigma''} \equiv \int d^3\vec{R}\,\psi^*_{\mathrm{m\kappa\gamma}}\psi_{\mathrm{m''\kappa''\gamma''}}\psi_{\mathrm{m'\kappa'\gamma'}} = 4\pi^2\delta_{\mathrm{m,(m'+m'')}}\delta(\kappa'+\kappa''-\kappa)\mathfrak{D}^{\gamma\ \gamma'\ \gamma''}_{\mathrm{mm'm''}}$$

$$= \delta_{\mathrm{m,(m'+m'')}}\delta(\kappa'+\kappa''-\kappa)\dfrac{2\pi^2}{\sqrt{\gamma\gamma'}}$$

$$\times\left\{\begin{array}{l}\delta(\gamma''-(\gamma-\gamma'))\mathrm{H}(\gamma-\gamma')\\ +(-1)^{\mathrm{m'}}\delta(\gamma''-(\gamma'+\gamma))+(-1)^{\mathrm{m-m'}}\delta(\gamma''-(\gamma'-\gamma))\mathrm{H}(\gamma'-\gamma)\end{array}\right\}$$

A7

$$\mathbb{P}^{(8)mm'm''ss's''}_{\sigma\sigma'\sigma''} \equiv \int d^3\vec{R}\, \vec{\chi}^*_{m\kappa\gamma s} \cdot \left(\vec{\chi}_{m'\kappa'\gamma's'} \cdot \vec{\nabla}\right) \vec{\chi}_{m''\kappa''\gamma''s''}$$

$$= 4\pi^2 i k^{-2} k'^{-2} k''^{-2} \delta_{m,(m'+m'')} \delta(\kappa' + \kappa'' - \kappa)$$

$$\times \begin{Bmatrix} \dfrac{\gamma\gamma'\gamma''^2}{4} (sk+\kappa)(s'k'+\kappa')(s''k''+\kappa'') \mathfrak{D}^{\gamma\ \gamma'\ \gamma''}_{(m+1)(m'+1)m''} \\[4pt] - \dfrac{\gamma\gamma'\gamma''^2}{4} (sk-\kappa)(s'k'-\kappa')(s''k''-\kappa'') \mathfrak{D}^{\gamma\ \gamma'\ \gamma''}_{(m-1)(m'-1)m''} \\[4pt] - \dfrac{\gamma\gamma'\gamma''^2}{4} (sk+\kappa)(s'k'-\kappa')(s''k''+\kappa'') \mathfrak{D}^{\gamma\ \gamma'\ \gamma''}_{(m+1)(m'-1)(m''+2)} \\[4pt] + \dfrac{\gamma\gamma'\gamma''^2}{4} (sk-\kappa)(s'k'+\kappa')(s''k''-\kappa'') \mathfrak{D}^{\gamma\ \gamma'\ \gamma''}_{(m-1)(m'+1)(m''-2)} \\[4pt] - \dfrac{\gamma}{2} \kappa''\gamma''\gamma'^2 (sk+\kappa)(s''k''+\kappa'') \mathfrak{D}^{\gamma\ \gamma'\ \gamma''}_{(m+1)m'(m''+1)} \\[4pt] - \dfrac{\gamma}{2} \kappa''\gamma''\gamma'^2 (sk-\kappa)(s''k''-\kappa'') \mathfrak{D}^{\gamma\ \gamma'\ \gamma''}_{(m-1)m'(m''-1)} \\[4pt] + \dfrac{\gamma^2\gamma'\gamma''^3}{2} (s'k'+\kappa') \mathfrak{D}^{\gamma\ \gamma'\ \gamma''}_{m(m'+1)(m''-1)} \\[4pt] - \dfrac{\gamma^2\gamma'\gamma''^3}{2} (s'k'-\kappa') \mathfrak{D}^{\gamma\ \gamma'\ \gamma''}_{m(m'-1)(m''+1)} - \gamma^2\gamma'^2 \kappa''\gamma''^2 \mathfrak{D}^{\gamma\ \gamma'\ \gamma''}_{mm'm''} \end{Bmatrix}$$

$$= 2\pi^2 i k^{-2} k'^{-2} k''^{-2} \sqrt{\gamma\gamma'}\, \delta_{m,(m'+m'')} \delta(\kappa' + \kappa'' - \kappa)$$

$$\times \begin{Bmatrix} \left\{(ss''kk'' + \kappa\kappa'' + \gamma\gamma'')(\kappa'\gamma'' - \kappa''\gamma') - \dfrac{1}{2} s's''k'k''\kappa\gamma'\right\} \gamma'' \delta(\gamma'' - (\gamma-\gamma')) H(\gamma-\gamma') \\[4pt] -(-1)^{m'} \left\{(ss''kk'' + \kappa\kappa'' + \gamma\gamma'')(\kappa'\gamma'' + \gamma'\kappa'') - \dfrac{1}{2} s's''k'k''\kappa\gamma''\right\} \gamma'' \delta(\gamma'' - (\gamma'+\gamma)) \\[4pt] +(-1)^{m-m'} \left\{(ss''kk'' + \kappa\kappa'' - \gamma\gamma'')(\kappa'\gamma'' + \kappa''\gamma') - \dfrac{1}{2} s'k's''k''\kappa\gamma\right\} \gamma'' \delta(\gamma'' - (\gamma'-\gamma)) H(\gamma'-\gamma) \end{Bmatrix}$$

A8

$$\mathbb{P}^{(9)mm'm''ss'}_{\sigma\sigma'\sigma''} \equiv \int d^3\vec{R}\,\vec{\chi}^*_{m\kappa\gamma s} \cdot \left(\vec{\chi}_{m'\kappa'\gamma's'} \cdot \vec{\nabla}\right)\vec{\nabla}\psi_{m''\kappa''\gamma''}$$

$$= 4\pi^2 k^{-2}k'^{-2}\delta_{m,(m'+m'')}\delta(\kappa'+\kappa''-\kappa)$$

$$\times \left\{\begin{array}{l} +\dfrac{1}{4}\gamma\gamma'\gamma''^2(sk-\kappa)(s'k'+\kappa')\mathfrak{D}^{\gamma\ \gamma'\ \gamma''}_{(m-1)(m'+1)(m''-2)} \\[4pt] -\dfrac{1}{4}\gamma\gamma'\gamma''^2(sk+\kappa)(s'k'+\kappa')\mathfrak{D}^{\gamma\ \gamma'\ \gamma''}_{(m+1)(m'+1)m''} \\[4pt] -\dfrac{1}{4}\gamma\gamma'\gamma''^2(sk-\kappa)(s'k'-\kappa')\mathfrak{D}^{\gamma\ \gamma'\ \gamma''}_{(m-1)(m'-1)m''} \\[4pt] +\dfrac{1}{4}\gamma\gamma'\gamma''^2(sk+\kappa)(s'k'-\kappa')\mathfrak{D}^{\gamma\ \gamma'\ \gamma''}_{(m+1)(m'-1)(m''+2)} \\[4pt] -\dfrac{1}{2}\gamma\gamma'^2\kappa''\gamma''(sk-\kappa)\mathfrak{D}^{\gamma\ \gamma'\ \gamma''}_{(m-1)m'(m''-1)} \\[4pt] +\dfrac{1}{2}\gamma\gamma'^2\kappa''\gamma''(sk+\kappa)\mathfrak{D}^{\gamma\ \gamma'\ \gamma''}_{(m+1)m'(m''+1)} \\[4pt] +\dfrac{1}{2}\gamma^2\gamma'\kappa''\gamma''(s'k'+\kappa')\mathfrak{D}^{\gamma\ \gamma'\ \gamma''}_{m(m'+1)(m''-1)} \\[4pt] -\dfrac{1}{2}\gamma^2\gamma'\kappa''\gamma''(s'k'-\kappa')\mathfrak{D}^{\gamma\ \gamma'\ \gamma''}_{m(m'-1)(m''+1)} - \gamma^2\kappa''^2\gamma'^2\mathfrak{D}^{\gamma\ \gamma'\ \gamma''}_{mm'm''} \end{array}\right\}$$

$$= 2\pi^2 k^{-2}k'^{-2}\sqrt{\gamma\gamma'}\,\delta_{m,(m'+m'')}\delta(\kappa'+\kappa''-\kappa) \quad\quad \text{A9}$$

$$\times \left\{\begin{array}{l} +(\gamma\kappa''-\gamma''\kappa)(\gamma''\kappa'-\gamma'\kappa'')\delta(\gamma''-(\gamma-\gamma'))H(\gamma-\gamma') \\[4pt] -(\gamma''\kappa'+\gamma'\kappa'')(\gamma\kappa''+\gamma''\kappa)(-1)^{m-m'}\delta(\gamma''-(\gamma'-\gamma))H(\gamma'-\gamma) \\[4pt] +(\kappa'\gamma''+\gamma'\kappa'')(\gamma''\kappa-\gamma\kappa'')(-1)^{m'}\delta(\gamma''-(\gamma'+\gamma)) \end{array}\right\}$$

$$\mathbb{P}^{(10)mm'm''ss''}_{\sigma\sigma'\sigma''} \equiv \int d^3\vec{R}\, \vec{\chi}^*_{m\kappa\gamma s} \cdot \left(\vec{\nabla}\psi_{m'\kappa'\gamma'} \cdot \vec{\nabla}\right) \vec{\chi}_{m''\kappa''\gamma''s''}$$

$$= 4\pi^2 k^{-2} k''^{-2} \delta_{m,(m'+m'')} \delta(\kappa' + \kappa'' - \kappa)$$

$$\times \left\{ \begin{array}{l} -\dfrac{\gamma\gamma'\gamma''^2}{4}(sk-\kappa)(s''k''-\kappa'')\left(\mathfrak{D}^{\gamma\;\gamma'\;\gamma''}_{(m-1)(m'-1)m''} + \mathfrak{D}^{\gamma\;\gamma'\;\gamma''}_{(m-1)(m'+1)(m''-2)}\right) \\ -\dfrac{\gamma\gamma'\gamma''^2}{4}(sk+\kappa)(s''k''+\kappa'')\left(\mathfrak{D}^{\gamma\;\gamma'\;\gamma''}_{(m+1)(m'-1)(m''+2)} + \mathfrak{D}^{\gamma\;\gamma'\;\gamma''}_{(m+1)(m'+1)m''}\right) \\ -\kappa'\kappa''\dfrac{\gamma\gamma''}{2}(sk+\kappa)(s''k''+\kappa'')\mathfrak{D}^{\gamma\;\gamma'\;\gamma''}_{(m+1)m'(m''+1)} \\ -\kappa'\kappa''\dfrac{\gamma\gamma''}{2}(sk-\kappa)(s''k''-\kappa'')\mathfrak{D}^{\gamma\;\gamma'\;\gamma''}_{(m-1)m'(m''-1)} \\ -\dfrac{1}{2}\gamma^2\gamma'\gamma''^3 \mathfrak{D}^{\gamma\;\gamma'\;\gamma''}_{m(m'-1)(m''+1)} + \dfrac{1}{2}\gamma^2\gamma'\gamma''^3 \mathfrak{D}^{\gamma\;\gamma'\;\gamma''}_{m(m'+1)(m''-1)} - \kappa''\kappa'\gamma^2\gamma''^2 \mathfrak{D}^{\gamma\;\gamma'\;\gamma''}_{mm'm''} \end{array} \right\}$$

$$= k^{-2} k''^{-2} \times \dfrac{2\pi^2}{\sqrt{\gamma\gamma'}} \delta_{m,(m'+m'')} \delta(\kappa'+\kappa''-\kappa)$$

$$\times \left\{ \begin{array}{l} -\left((\gamma'\gamma''+\kappa'\kappa'')(ss''kk''+\kappa\kappa'') + \kappa'\kappa''\gamma\gamma''\right)\gamma\gamma''\delta(\gamma''-(\gamma-\gamma'))H(\gamma-\gamma') \\ -\left((\gamma'\gamma''-\kappa'\kappa'')(ss''kk''+\kappa\kappa'') + \kappa''\kappa\gamma\gamma''\right)(-1)^{m-m'}\gamma\gamma''\delta(\gamma''-(\gamma'-\gamma))H(\gamma'-\gamma) \\ +\left((\gamma'\gamma''-\kappa'\kappa'')(ss''kk''+\kappa\kappa'') - \kappa''\kappa'\gamma\gamma''\right)(-1)^{m'}\gamma\gamma''\delta(\gamma''-(\gamma'+\gamma)) \end{array} \right\} \quad \text{A10}$$

$$\mathbb{P}^{(11)mm'm''s}_{\sigma\sigma'\sigma''} = \int d^3\vec{R}\, \vec{\chi}^*_{m\kappa\gamma s} \cdot \left(\vec{\nabla}\psi_{m'\kappa'\gamma'} \cdot \vec{\nabla}\right) \vec{\nabla}\psi_{m''\kappa''\gamma''}$$

$$= 4\pi^2 i k^{-2} \delta_{m,(m'+m'')} \delta(\kappa' + \kappa'' - \kappa)$$

$$\times \begin{Bmatrix} \dfrac{1}{4}\gamma\gamma'\gamma''^2 (sk-\kappa)\left(\mathfrak{D}^{\gamma\;\;\gamma'\;\;\gamma''}_{(m-1)(m'+1)(m''-2)} + \mathfrak{D}^{\gamma\;\;\gamma'\;\;\gamma''}_{(m-1)(m'-1)m''}\right) \\ -\dfrac{1}{4}\gamma\gamma'\gamma''^2 (sk+\kappa)\left(\mathfrak{D}^{\gamma\;\;\gamma'\;\;\gamma''}_{(m+1)(m'+1)m''} + \mathfrak{D}^{\gamma\;\;\gamma'\;\;\gamma''}_{(m+1)(m'-1)(m''+2)}\right) \\ +\dfrac{1}{2}\gamma\kappa''\kappa'\gamma''(sk-\kappa)\mathfrak{D}^{\gamma\;\;\gamma'\;\;\gamma''}_{(m-1)m'(m''-1)} - \dfrac{1}{2}\gamma\kappa''\kappa'\gamma''(sk+\kappa)\mathfrak{D}^{\gamma\;\;\gamma'\;\;\gamma''}_{(m+1)m'(m''+1)} \\ +\gamma^2\kappa''\dfrac{\gamma'\gamma''}{2}\mathfrak{D}^{\gamma\;\;\gamma'\;\;\gamma''}_{m(m'+1)(m''-1)} + \gamma^2\kappa''\dfrac{\gamma'\gamma''}{2}\mathfrak{D}^{\gamma\;\;\gamma'\;\;\gamma''}_{m(m'-1)(m''+1)} + \gamma^2\kappa'\kappa''\kappa''\mathfrak{D}^{\gamma\;\gamma'\;\gamma''}_{mm'm''} \end{Bmatrix}$$

$$= 2\pi^2 i k^{-2} \sqrt{\gamma\gamma'^{-1}} \delta_{m,(m'+m'')} \delta(\kappa'+\kappa''-\kappa)$$

$$\times \begin{Bmatrix} +\left(\kappa'\kappa''+\gamma'\gamma''\right)\left(\gamma\kappa''-\kappa\gamma''\right)\delta\left(\gamma''-(\gamma-\gamma')\right)H(\gamma-\gamma') \\ +\left(\gamma''\kappa+\gamma\kappa''\right)\left(\kappa'\kappa''-\gamma'\gamma''\right)(-1)^{m-m'}\delta\left(\gamma''-(\gamma'-\gamma)\right)H(\gamma'-\gamma) \\ +\left(\gamma\kappa''-\gamma''\kappa\right)\left(\kappa''\kappa'-\gamma'\gamma''\right)(-1)^{m'}\delta\left(\gamma''-(\gamma'+\gamma)\right) \end{Bmatrix} \quad \text{A11}$$

$$\mathbb{P}^{(12)mm'm''ss's''}_{\sigma\sigma'\sigma''} \equiv \int d^3\vec{R}\,\vec{\chi}^*_{m\kappa\gamma s} \cdot \vec{\chi}_{m'\kappa'\gamma's'} \times \vec{\chi}_{m''\kappa''\gamma''s''}$$

$$= 4\pi^2 ik^{-2}k'^{-2}k''^{-2}\delta_{m,(m'+m'')}\delta(\kappa' + \kappa'' - \kappa)$$

$$\times \begin{Bmatrix} +\dfrac{1}{2}\gamma\gamma'^2\gamma''(sk+\kappa)(s''k''+\kappa'')\mathfrak{D}^{\gamma\ \gamma'\ \gamma''}_{(m+1)m'(m''+1)} \\ -\dfrac{1}{2}\gamma\gamma'^2\gamma''(sk+\kappa)(s'k'+\kappa')\mathfrak{D}^{\gamma\ \gamma'\ \gamma''}_{(m+1)(m'+1)m''} \\ +\dfrac{1}{2}\gamma\gamma'^2\gamma''(sk-\kappa)(s'k'-\kappa')\mathfrak{D}^{\gamma\ \gamma'\ \gamma''}_{(m-1)(m'-1)m''} \\ -\dfrac{1}{2}\gamma\gamma'^2\gamma''(sk-\kappa)(s''k''-\kappa'')\mathfrak{D}^{\gamma\ \gamma'\ \gamma''}_{(m-1)m'(m''-1)} \\ +\dfrac{1}{2}\gamma'\gamma''\gamma^2(s'k'-\kappa')(s''k''+\kappa'')\mathfrak{D}^{\gamma\ \gamma'\ \gamma''}_{m(m'-1)(m''+1)} \\ -\dfrac{1}{2}\gamma'\gamma''\gamma^2(s'k'+\kappa')(s''k''-\kappa'')\mathfrak{D}^{\gamma\ \gamma'\ \gamma''}_{m(m'+1)(m''-1)} \end{Bmatrix}$$

$$= 2\pi^2 ik^{-2}k'^{-2}k''^{-2}\sqrt{\gamma\gamma'}\,\delta_{m,(m'+m'')}\delta(\kappa'+\kappa''-\kappa)$$

$$\times \left\{ \begin{aligned} &\begin{bmatrix} \gamma'(sk\kappa''+s''k''\kappa) \\ -\gamma'(sk\kappa'+s'k'\kappa) \\ +\gamma(s'k'\kappa''-s''k''\kappa') \end{bmatrix} \gamma''\delta(\gamma''-(\gamma-\gamma'))H(\gamma-\gamma') \\ &-\begin{bmatrix} \gamma'(sk\kappa''+s''k''\kappa) \\ +\gamma'(sk\kappa'+s'k'\kappa) \\ +\gamma(s'k'\kappa''-s''k''\kappa') \end{bmatrix}(-1)^{m-m'}\gamma''\delta(\gamma''-(\gamma'-\gamma))H(\gamma'-\gamma) \\ &+\begin{bmatrix} \gamma'(sk\kappa''+s''k''\kappa) \\ +\gamma'(sk\kappa'+s'k'\kappa) \\ -\gamma(s'k'\kappa''-s''k''\kappa') \end{bmatrix}(-1)^{m'}\gamma''\delta(\gamma''-(\gamma'+\gamma)) \end{aligned} \right\} \quad \text{A12}$$

$$\mathbb{P}^{(13)\mathrm{mm'm''ss''}}_{\sigma\sigma'\sigma''} \equiv \int d^3\vec{R}\, \vec{\chi}^*_{m\kappa\gamma s} \cdot \vec{\nabla}\psi_{m'\kappa'\gamma'} \times \vec{\chi}_{m''\kappa''\gamma''s''}$$

$$= 4\pi^2 \delta_{m,(m'+m'')} k^{-2} k''^{-2} \delta(\kappa' + \kappa'' - \kappa)$$

$$\times \begin{Bmatrix} \dfrac{1}{2}\gamma\gamma'\gamma''^2 (sk - \kappa)\mathfrak{D}^{\gamma\ \gamma'\ \gamma''}_{(m-1)(m'-1)m''} \\[4pt] +\dfrac{1}{2}\gamma\gamma'\gamma''^2 (sk + \kappa)\mathfrak{D}^{\gamma\ \gamma'\ \gamma''}_{(m+1)(m'+1)m''} \\[4pt] +\dfrac{1}{2}\gamma\kappa'\gamma'' (sk + \kappa)(s''k'' + \kappa'')\mathfrak{D}^{\gamma\ \gamma'\ \gamma''}_{(m+1)m'(m''+1)} \\[4pt] -\dfrac{1}{2}\gamma\kappa'\gamma'' (sk - \kappa)(s''k'' - \kappa'')\mathfrak{D}^{\gamma\ \gamma'\ \gamma''}_{(m-1)m'(m''-1)} \\[4pt] +\dfrac{1}{2}\gamma'\gamma''\gamma^2 (s''k'' + \kappa'')\mathfrak{D}^{\gamma\ \gamma'\ \gamma''}_{m(m'-1)(m''+1)} \\[4pt] +\dfrac{1}{2}\gamma'\gamma''\gamma^2 (s''k'' - \kappa'')\mathfrak{D}^{\gamma\ \gamma'\ \gamma''}_{m(m'+1)(m''-1)} \end{Bmatrix}$$

$$= \delta_{m,(m'+m'')} \frac{2\pi^2}{\sqrt{\gamma\gamma'}} k^{-2} k''^{-2} \delta(\kappa' + \kappa'' - \kappa)$$

$$\times \begin{Bmatrix} \left[ sk(\gamma'\gamma'' + \kappa'\kappa'') + s''k''(\gamma\gamma' + \kappa\kappa') \right]\gamma\gamma''\delta(\gamma'' - (\gamma - \gamma'))\mathrm{H}(\gamma - \gamma') \\ +\left[ sk\gamma\gamma''(\gamma'\gamma'' - \kappa'\kappa'') - s''k''\gamma\gamma''(\gamma\gamma' + \kappa\kappa') \right](-1)^{m-m'}\delta(\gamma'' - (\gamma' - \gamma))\mathrm{H}(\gamma' - \gamma) \\ -\left[ sk\gamma\gamma''(\gamma'\gamma'' - \kappa'\kappa'') + s''k''\gamma\gamma''(\gamma\gamma' - \kappa\kappa') \right](-1)^{m'}\delta(\gamma'' - (\gamma' + \gamma)) \end{Bmatrix}$$

A13

$$\mathbb{P}^{(14)mm'm''s's''}_{\sigma\sigma'\sigma''} \equiv \int d^3\vec{R}\, \vec{\nabla}\psi^*_{m\kappa\gamma} \cdot \left(\vec{\chi}_{m'\kappa'\gamma's'} \cdot \vec{\nabla}\right) \vec{\chi}_{m''\kappa''\gamma''s''}$$

$$= 4\pi^2 \delta_{m,(m'+m'')} \delta(\kappa' + \kappa'' - \kappa) k''^{-2} k'^{-2}$$

$$\times \begin{cases} \dfrac{1}{4}\gamma\gamma'\gamma''^2 (s'k'+\kappa')(s''k''+\kappa'') \mathfrak{D}^{\gamma\phantom{'}\gamma'\phantom{''}\gamma''}_{(m+1)(m'+1)m''} \\[4pt] -\dfrac{1}{4}\gamma\gamma'\gamma''^2 (s'k'+\kappa')(s''k''-\kappa'') \mathfrak{D}^{\gamma\phantom{'}\gamma'\phantom{''}\gamma''}_{(m-1)(m'+1)(m''-2)} \\[4pt] +\dfrac{1}{4}\gamma\gamma'\gamma''^2 (s'k'-\kappa')(s''k''-\kappa'') \mathfrak{D}^{\gamma\phantom{'}\gamma'\phantom{''}\gamma''}_{(m-1)(m'-1)m''} \\[4pt] -\dfrac{1}{4}\gamma\gamma'\gamma''^2 (s'k'-\kappa')(s''k''+\kappa'') \mathfrak{D}^{\gamma\phantom{'}\gamma'\phantom{''}\gamma''}_{(m+1)(m'-1)(m''+2)} \\[4pt] -\dfrac{\gamma}{2}\kappa''\gamma''\gamma'^2 (s''k''+\kappa'') \mathfrak{D}^{\gamma\phantom{'}\gamma'\phantom{''}\gamma''}_{(m+1)m'(m''+1)} \\[4pt] +\dfrac{\gamma}{2}\kappa''\gamma''\gamma'^2 (s''k''-\kappa'') \mathfrak{D}^{\gamma\phantom{'}\gamma'\phantom{''}\gamma''}_{(m-1)m'(m''-1)} \\[4pt] -\dfrac{\kappa\gamma'\gamma''^3}{2}(s'k'+\kappa') \mathfrak{D}^{\gamma\phantom{'}\gamma'\phantom{''}\gamma''}_{m(m'+1)(m''-1)} \\[4pt] -\dfrac{\kappa\gamma'\gamma''^3}{2}(s'k'-\kappa') \mathfrak{D}^{\gamma\phantom{'}\gamma'\phantom{''}\gamma''}_{m(m'-1)(m''+1)} - \kappa\gamma'^2\kappa''\gamma''^2 \mathfrak{D}^{\gamma\,\gamma'\,\gamma''}_{mm'm''} \end{cases}$$

$$= 2\pi^2 \delta_{m,(m'+m'')} \sqrt{\gamma^{-1}\gamma'}\, \delta(\kappa'+\kappa''-\kappa) k''^{-2} k'^{-2}$$

$$\times \begin{cases} \left[\gamma\kappa''(\kappa'\gamma''-\gamma'\kappa'') - \kappa\gamma''(s'k'\gamma''+\gamma'\kappa'')\right]\gamma''\delta(\gamma''-(\gamma-\gamma'))H(\gamma-\gamma') \\[4pt] +\left[\gamma\kappa''(\kappa'\gamma''+\gamma'\kappa'') + \kappa\gamma''(s'k'\gamma''-\gamma'\kappa'')\right](-1)^{m-m'}\gamma''\delta(\gamma''-(\gamma'-\gamma))H(\gamma'-\gamma) \\[4pt] +\left[-\gamma\kappa''(\gamma''\kappa'+\kappa''\gamma') + \kappa\gamma''(s'k'\gamma''-\gamma'\kappa'')\right](-1)^{m'}\gamma''\delta(\gamma''-(\gamma'+\gamma)) \end{cases}$$

$$\text{A14}$$

$$\mathbb{P}^{(15)mm'm''s'}_{\sigma\sigma'\sigma''} = \int d^3\vec{R}\vec{\nabla}\psi^*_{m\kappa\gamma} \cdot \left(\vec{\chi}_{m'\kappa'\gamma's'} \cdot \vec{\nabla}\right)\vec{\nabla}\psi_{m''\kappa''\gamma''}$$

$$= 4\pi^2 ik'^{-2}\delta_{m,(m'+m'')}\delta(\kappa'+\kappa''-\kappa)$$

$$\times \begin{cases} \dfrac{1}{4}\gamma\gamma'\gamma''^2(s'k'+\kappa')\mathfrak{D}^{\gamma\quad\gamma'\quad\gamma''}_{(m-1)(m'+1)(m''-2)} + \dfrac{1}{4}\gamma\gamma'\gamma''^2(s'k'+\kappa')\mathfrak{D}^{\gamma\quad\gamma'\quad\gamma''}_{(m+1)(m'+1)m''} \\ -\dfrac{1}{4}\gamma\gamma'\gamma''^2(s'k'-\kappa')\mathfrak{D}^{\gamma\quad\gamma'\quad\gamma''}_{(m-1)(m'-1)m''} - \dfrac{1}{4}\gamma\gamma'\gamma''^2(s'k'-\kappa')\mathfrak{D}^{\gamma\quad\gamma'\quad\gamma''}_{(m+1)(m'-1)(m''+2)} \\ +\kappa\kappa''\dfrac{\gamma'\gamma''}{2}(s'k'+\kappa')\mathfrak{D}^{\gamma\quad\gamma'\quad\gamma''}_{m(m'+1)(m''-1)} - \kappa\kappa''\dfrac{\gamma'\gamma''}{2}(s'k'-\kappa')\mathfrak{D}^{\gamma\quad\gamma'\quad\gamma''}_{m(m'-1)(m''+1)} \\ -\gamma\kappa''\gamma'^2\dfrac{\gamma''}{2}\mathfrak{D}^{\gamma\quad\gamma'\quad\gamma''}_{(m-1)m'(m''-1)} - \dfrac{1}{2}\gamma\kappa''\gamma'^2\gamma''\mathfrak{D}^{\gamma\quad\gamma'\quad\gamma''}_{(m+1)m'(m''+1)} - \kappa\kappa''^2\gamma'^2\mathfrak{D}^{\gamma\gamma'\gamma''}_{mm'm''} \end{cases}$$

$$= 2\pi^2 ik'^{-2}\delta_{m,(m'+m'')}\sqrt{\gamma^{-1}\gamma'}\delta(\kappa'+\kappa''-\kappa)$$

$$\times \begin{cases} (\gamma\gamma''+\kappa\kappa'')(\kappa'\gamma''-\kappa''\gamma')\delta(\gamma''-(\gamma-\gamma'))H(\gamma-\gamma') \\ +(\kappa'\gamma''+\kappa''\gamma')(\gamma\gamma''-\kappa\kappa'')(-1)^{m-m'}\delta(\gamma''-(\gamma'-\gamma))H(\gamma'-\gamma) \\ -(\kappa'\gamma''+\gamma'\kappa'')(\gamma\gamma''+\kappa\kappa'')(-1)^{m'}\delta(\gamma''-(\gamma'+\gamma)) \end{cases} \qquad \text{A15}$$

$$\mathbb{P}^{(16)mm'm''s''}_{\sigma\sigma'\sigma''} = \int d^3\vec{R}\,\vec{\nabla}\psi^*_{m\kappa\gamma} \cdot \left(\vec{\nabla}\psi_{m'\kappa'\gamma'} \cdot \vec{\nabla}\right)\vec{\chi}_{m''\kappa''\gamma''s''}$$

$$= 4\pi^2 \frac{1}{2} i\gamma''k''^{-2}\delta_{m,(m'+m'')}\delta\left(\kappa'+\kappa''-\kappa\right)$$

$$\times \begin{Bmatrix} -\dfrac{1}{2}\gamma\gamma'\gamma''\left(s''k''-\kappa''\right)\mathfrak{D}^{\gamma\ \gamma'\ \gamma''}_{(m-1)(m'-1)m''} + \dfrac{1}{2}\gamma\gamma'\gamma''\left(s''k''+\kappa''\right)\mathfrak{D}^{\gamma\ \gamma'\ \gamma''}_{(m+1)(m'+1)m''} \\ +\dfrac{1}{2}\gamma\gamma'\gamma''\left(s''k''+\kappa''\right)\mathfrak{D}^{\gamma\ \gamma'\ \gamma''}_{(m+1)(m'-1)(m''+2)} - \dfrac{1}{2}\gamma\gamma'\gamma''\left(s''k''-\kappa''\right)\mathfrak{D}^{\gamma\ \gamma'\ \gamma''}_{(m-1)(m'+1)(m''-2)} \\ +\gamma\kappa'\kappa''\left(s''k''+\kappa''\right)\mathfrak{D}^{\gamma\ \gamma'\ \gamma''}_{(m+1)m'(m''+1)} - \gamma\kappa'\kappa''\left(s''k''-\kappa''\right)\mathfrak{D}^{\gamma\ \gamma'\ \gamma''}_{(m-1)m'(m''-1)} \\ +\dfrac{1}{2}\kappa\gamma'\gamma''^2\mathfrak{D}^{\gamma\ \gamma'\ \gamma''}_{m(m'-1)(m''+1)} + \dfrac{1}{2}\kappa\gamma'\gamma''^2\mathfrak{D}^{\gamma\ \gamma'\ \gamma''}_{m(m'+1)(m''-1)} + \kappa\kappa''\kappa'\gamma''\mathfrak{D}^{\gamma\ \gamma'\ \gamma''}_{mm'm''} \end{Bmatrix}$$

$$= i\gamma''k''^{-2}\delta_{m,(m'+m'')}\frac{\pi^2}{\sqrt{\gamma\gamma'}}\delta\left(\kappa'+\kappa''-\kappa\right)$$

$$\times \begin{Bmatrix} +\left(\gamma'\gamma''+\kappa'\kappa''\right)\left(2\gamma\kappa''+\kappa\gamma''\right)\delta\left(\gamma''-(\gamma-\gamma')\right)\mathrm{H}(\gamma-\gamma') \\ +\left(\gamma'\gamma''-\kappa'\kappa''\right)\left(2\gamma\kappa''-\kappa\gamma''\right)(-1)^{m-m'}\delta\left(\gamma''-(\gamma'-\gamma)\right)\mathrm{H}(\gamma'-\gamma) \\ +\left(\kappa'\kappa''-\gamma'\gamma''\right)\left(2\gamma\kappa''-\kappa\gamma''\right)(-1)^{m'}\delta\left(\gamma''-(\gamma'+\gamma)\right) \end{Bmatrix} \qquad \text{A16}$$

$$\mathbb{P}^{(17)\mathrm{mm'm''}}_{\sigma\sigma'\sigma''} = \int d^3\vec{R}\,\vec{\nabla}\psi^*_{\mathrm{m\kappa\gamma}} \cdot \left(\vec{\nabla}\psi_{\mathrm{m'\kappa'\gamma'}} \cdot \vec{\nabla}\right)\vec{\nabla}\psi_{\mathrm{m''\kappa''\gamma''}}$$

$$= -4\pi^2 \delta_{\mathrm{m,(m'+m'')}}\delta(\kappa'+\kappa''-\kappa)$$

$$\times \left\{\begin{array}{l} \dfrac{1}{4}\gamma\gamma'\gamma''^2\left(\mathfrak{D}^{\gamma\;\;\gamma'\;\;\gamma''}_{(\mathrm{m}-1)(\mathrm{m'}+1)(\mathrm{m''}-2)} + \mathfrak{D}^{\gamma\;\;\gamma'\;\;\gamma''}_{(\mathrm{m}-1)(\mathrm{m'}-1)\mathrm{m''}} + \mathfrak{D}^{\gamma\;\;\gamma'\;\;\gamma''}_{(\mathrm{m}+1)(\mathrm{m'}+1)\mathrm{m''}} + \mathfrak{D}^{\gamma\;\;\gamma'\;\;\gamma''}_{(\mathrm{m}+1)(\mathrm{m'}-1)(\mathrm{m''}+2)}\right) \\[6pt] +\dfrac{1}{2}\kappa''\kappa'\gamma\gamma''\left(\mathfrak{D}^{\gamma\;\;\gamma'\;\;\gamma''}_{(\mathrm{m}-1)\mathrm{m'}(\mathrm{m''}-1)} + \mathfrak{D}^{\gamma\;\;\gamma'\;\;\gamma''}_{(\mathrm{m}+1)\mathrm{m'}(\mathrm{m''}+1)}\right) + \kappa\kappa'\kappa''^2\mathfrak{D}^{\gamma\,\gamma'\,\gamma''}_{\mathrm{mm'm''}} \\[6pt] +\dfrac{1}{2}\kappa\kappa''\gamma'\gamma''\left(\mathfrak{D}^{\gamma\;\;\gamma'\;\;\gamma''}_{\mathrm{m(m'+1)(m''-1)}} + \mathfrak{D}^{\gamma\;\;\gamma'\;\;\gamma''}_{\mathrm{m(m'-1)(m''+1)}}\right) \end{array}\right\}$$

$$= -\delta_{\mathrm{m,(m'+m'')}}\frac{2\pi^2}{\sqrt{\gamma\gamma'}}\delta(\kappa'+\kappa''-\kappa)$$

$$\times \left\{\begin{array}{l} (\gamma\gamma''+\kappa\kappa'')(\gamma'\gamma''+\kappa''\kappa')\delta(\gamma''-(\gamma-\gamma'))\mathrm{H}(\gamma-\gamma') \\[4pt] +(-1)^{\mathrm{m-m'}}(\gamma\gamma''-\kappa\kappa'')(\gamma'\gamma''-\kappa''\kappa')\delta(\gamma''-(\gamma'-\gamma))\mathrm{H}(\gamma'-\gamma) \\[4pt] -(-1)^{\mathrm{m'}}(\gamma\gamma''+\kappa\kappa'')(\gamma'\gamma''-\kappa''\kappa')\delta(\gamma''-(\gamma'+\gamma)) \end{array}\right\} \qquad \mathrm{A17}$$

$$\mathbb{P}^{(18)mm'm''s's''}_{\sigma\sigma'\sigma''} = \int d^3\vec{R}\vec{\nabla}\psi^*_{m\kappa\gamma} \cdot \vec{\chi}_{m'\kappa'\gamma's'} \times \vec{\chi}_{m''\kappa''\gamma''s''}$$

$$= 4\pi^2\delta_{m,(m'+m'')}\delta(\kappa'+\kappa''-\kappa)k'^{-2}k''^{-2}$$

$$\times \begin{Bmatrix} -\dfrac{1}{2}\gamma\gamma'\gamma''^2(s'k'+\kappa')\mathfrak{D}^{\gamma\quad\gamma'\quad\gamma''}_{(m+1)(m'+1)m''} - \dfrac{1}{2}\gamma\gamma'\gamma''^2(s'k'-\kappa')\mathfrak{D}^{\gamma\quad\gamma'\quad\gamma''}_{(m-1)(m'-1)m''} \\ +\dfrac{1}{2}\gamma\gamma'^2\gamma''(s''k''+\kappa'')\mathfrak{D}^{\gamma\quad\gamma'\quad\gamma''}_{(m+1)m'(m''+1)} + \dfrac{1}{2}\gamma\gamma'^2\gamma''(s''k''-\kappa'')\mathfrak{D}^{\gamma\quad\gamma'\quad\gamma''}_{(m-1)m'(m''-1)} \\ -\dfrac{1}{2}\kappa\gamma'\gamma''(s'k'-\kappa')(s''k''+\kappa'')\mathfrak{D}^{\gamma\quad\gamma'\quad\gamma''}_{m(m'-1)(m''+1)} \\ +\dfrac{1}{2}\kappa\gamma'\gamma''(s'k'+\kappa')(s''k''-\kappa'')\mathfrak{D}^{\gamma\quad\gamma'\quad\gamma''}_{m(m'+1)(m''-1)} \end{Bmatrix}$$

$$= 2\pi^2\delta_{m,(m'+m'')}\sqrt{\gamma'\gamma^{-1}}\delta(\kappa'+\kappa''-\kappa)k'^{-2}k''^{-2}$$

$$\times \begin{Bmatrix} +\{s''k''(\gamma\gamma'+\kappa'\kappa) - s'k'(\gamma\gamma''+\kappa''\kappa)\}\gamma''\delta(\gamma''-(\gamma-\gamma'))H(\gamma-\gamma') \\ -\{s'k'(\gamma\gamma''-\kappa''\kappa) + s''k''(\kappa'\kappa+\gamma\gamma')\}(-1)^{m-m'}\gamma''\delta(\gamma''-(\gamma'-\gamma))H(\gamma'-\gamma) \\ +(s'k'(\gamma\gamma''+\kappa\kappa'') + s''k''(\gamma\gamma'-\kappa\kappa'))(-1)^{m'}\gamma''\delta(\gamma''-(\gamma'+\gamma)) \end{Bmatrix} \quad \text{A18}$$

$$\mathbb{P}^{(19)mm'm''s''}_{\sigma\sigma'\sigma''} \equiv \int d^3\vec{R}\,\vec{\nabla}\psi^*_{m\kappa\gamma} \cdot \left(\vec{\nabla}\psi_{m'\kappa'\gamma'} \times \vec{\chi}_{m''\kappa''\gamma''s''}\right)$$

$$= 4\pi^2 i k''^{-2}\delta_{m,(m'+m'')}\delta(\kappa'+\kappa''-\kappa)$$

$$\times \left\{ \begin{array}{l} \dfrac{1}{2}\gamma\gamma'\gamma''^2 \left(\mathfrak{D}^{\gamma\ \gamma'\ \gamma''}_{(m-1)(m'-1)m''} - \mathfrak{D}^{\gamma\ \gamma'\ \gamma''}_{(m+1)(m'+1)m''}\right) \\ -\dfrac{1}{2}\gamma\gamma''\kappa'\left((s''k''+\kappa'')\mathfrak{D}^{\gamma\ \gamma'\ \gamma''}_{(m+1)m'(m''+1)} + (s''k''-\kappa'')\mathfrak{D}^{\gamma\ \gamma'\ \gamma''}_{(m-1)m'(m''-1)}\right) \\ +\dfrac{1}{2}\kappa\gamma'\gamma''\left((s''k''+\kappa'')\mathfrak{D}^{\gamma\ \gamma'\ \gamma''}_{m(m'-1)(m''+1)} + (s''k''-\kappa'')\mathfrak{D}^{\gamma\ \gamma'\ \gamma''}_{m(m'+1)(m''-1)}\right) \end{array} \right\}$$

$$= ik''^{-2}\delta_{m,(m'+m'')}\frac{2\pi^2}{\sqrt{\gamma\gamma'}}\delta(\kappa'+\kappa''-\kappa)$$

$$\times \left\{ \begin{array}{l} +s''k''(\kappa\gamma'-\gamma\kappa')\gamma''\delta(\gamma''-(\gamma-\gamma'))\,\mathrm{H}(\gamma-\gamma') \\ +s''k''(\gamma\kappa'-\kappa\gamma')(-1)^{m-m'}\gamma''\delta(\gamma''-(\gamma'-\gamma))\,\mathrm{H}(\gamma'-\gamma) \\ -s''k''(\gamma\kappa'+\kappa\gamma')(-1)^{m'}\gamma''\delta(\gamma''-(\gamma'+\gamma)) \end{array} \right\} \quad \text{A19}$$

Appendix II

Some results of the stochastic model

$$\left\langle \mathbb{R}_{ms}^{-1}\left(\vec{L},\kappa,\gamma\right)\right\rangle = \int_0^{2\pi} d\Theta \int_{-\infty}^{\infty} dL_z \int_0^{\infty} L_r dL_r p(\vec{L}) \mathbb{R}_{ms}^{-1}\left(\vec{L},\kappa,\gamma\right)$$

$$= \frac{(2\pi)^{-0.25} \delta_{m,0}}{k\sqrt{\gamma}} 2\sqrt{\frac{\pi}{3}} \Lambda^{-3.5} \exp\left(-\kappa^2 \Lambda^2\right) \int_0^{\infty} L_r dL_r \frac{1}{J_0(\gamma L_r)} \left(\exp\left(-3L_r^2/4\Lambda^2\right)\right) \qquad \text{A-20}$$

$$\to 0 \text{ as } \Lambda \to \infty$$

$$\left\langle \mathbb{R}_{ms}^{-1}\left(\vec{L},\kappa,\gamma\right) \mathbb{R}_{m''s''}\left(\vec{L},\kappa'',\gamma''\right) \mathbb{R}_{m's'}\left(\vec{L},\kappa',\gamma'\right)\right\rangle$$

$$= \sqrt{\frac{\gamma'\gamma''}{\gamma}} \frac{k'k''}{k} \frac{\delta_{s',s_0} \delta_{s'',s_0}}{\delta_{s,s_0}} (2\pi)^{-0.75} 2\sqrt{\pi} \delta_{m,(m'+m'')} \exp\left(-\left(\kappa'+\kappa''-\kappa\right)^2 \Lambda^2\right) \qquad \text{A-21}$$

$$\times \int_0^{\infty} L_r dL_r \frac{J_{m'}(\gamma' L_r) J_m(\gamma'' L_r)}{J_m(\gamma L_r)} \left\{\Lambda^{-0.5} \exp\left(-L_r^2/4\Lambda^2\right)\right\} \to 0 \text{ as } \Lambda \to \infty$$

$$\left\langle \mathbb{R}_{ms}^{-1}\left(\vec{L},\kappa,\gamma\right) \mathbb{R}_{m's'}\left(\vec{L},\kappa',\gamma'\right)\right\rangle = \int_0^{2\pi} d\Theta \int_{-\infty}^{\infty} dL_z \int_0^{\infty} L_r dL_r p(\vec{L}) \mathbb{R}_{ms}^{-1}\left(\vec{L},\kappa,\gamma\right) \mathbb{R}_{m's'}\left(\vec{L},\kappa',\gamma'\right)$$

$$= (2\pi)^{-0.5} \frac{\delta_{s',s_0} \sqrt{\gamma'} k'}{\delta_{s,s_0} \sqrt{\gamma} k} \Lambda^{-3} \delta_{m,m'} \int_{-\infty}^{\infty} dL_z \exp\left(-i(\kappa'-\kappa)L_z\right) \exp\left(-L_z^2/2\Lambda^2\right) \qquad \text{A-22}$$

$$\times \int_0^{\infty} L_r dL_r \exp\left(-L_r^2/2\Lambda^2\right) \frac{J_{m'}(\gamma' L_r)}{J_m(\gamma L_r)}$$

This value is equal to 1 for $\kappa' = \kappa; \gamma' = \gamma; s' = s$. For $s' \neq s$, choose $s_0 = s$, so that $\delta_{s',s_0} = 0$ and the expectation value given by A-22 becomes zero. For $\kappa' \neq \kappa; \gamma' = \gamma; s' = s$,

$$\int_{-\infty}^{\infty} dL_z \exp\left(-i(\kappa'-\kappa)L_z\right) \exp\left(-L_z^2/2\Lambda^2\right)$$

$$= \exp\left(-\tfrac{1}{2}(\kappa'-\kappa)^2 \Lambda^2\right) \to 0 \text{ as } \Lambda \to \infty$$

For $\kappa' = \kappa; \gamma' \neq \gamma, \gamma \neq 0; s' = s$,

$$\int_0^\infty L_r dL_r \frac{J_{m'}(\gamma' L_r)}{J_m(\gamma L_r)} \left\{ \lim_{\Lambda \to \infty} \Lambda^{-3} \exp\left(-L_r^2/2\Lambda^2\right) \right\} = 0$$

Hence, for $\kappa' = \kappa; \gamma' \neq \gamma, \gamma \neq 0; s' = s$, the expectation value in A-22 is zero in the limit of large $\Lambda$. For $\kappa' = \kappa; \gamma' \neq \gamma, \gamma = 0; s' = s$, the left hand side of A-22 is not defined. Since $\kappa, \kappa', \gamma, \gamma'$ are continuous variables and expectation values are to be defined as distributions, the above discussion can be summarized in the expression

$$\left\langle \mathbb{R}_{ms}^{-1}(\vec{L}, \kappa, \gamma) \mathbb{R}_{m's'}(\vec{L}, \kappa', \gamma') \right\rangle = \delta_{m,m'} \delta_{s,s'} \delta(\kappa - \kappa') \delta(\gamma - \gamma') \qquad \text{A-23}$$

## Appendix III

**Separation of random and non-random parts of spectral evolution equations**

Non-Random Parts:

The non-random part of spectral evolution equations 18-24 has two groups of terms. The first group contains products of non-random spectra. They have the same form as the terms on the right hand side of the original equation except that the spectral coefficients in the product have underbars. They are collectively denoted in this appendix by the symbol $[\![\_\_]\!]$. The contribution of a term involving product of random spectra with coefficient proportional to the function $\mathbb{P}^{(n)}$ to the non-random spectral evolution is denoted by $[\![\mathbb{P}^{(n)}_{\sim\sim}]\!]$

Equations 16-19: not written as they are trivial.

Equation 20: $\underline{\boldsymbol{J}}^S_{ms}(\sigma, t) = [\![\_\_]\!] + \sum_{\iota}\left([\![\mathbb{P}^{(1)}_{\sim\sim}]\!]_{\iota} + [\![\mathbb{P}^{(2)}_{\sim\sim}]\!]_{\iota}\right)$

$$[\![\mathbb{P}^{(1)}_{\sim\sim}]\!]_{\iota} = \frac{q_{\iota}ek^2}{8\sqrt{2}}\sqrt{\frac{\gamma+\gamma_0}{\gamma}}\delta_{m,0}\tfrac{1}{2}\{\delta(\kappa-\kappa_0)+\delta(\kappa+\kappa_0)\}\sum_{s'}\sum_{m'=-((\gamma+\gamma_0)\lambda_d/2)^{-1}}^{((\gamma+\gamma_0)\lambda_d/2)^{-1}}(-1)^{m'}$$

$$\int'\delta(\gamma'-(\gamma+\gamma_0)/2)\frac{\{ss'kk'+\kappa\kappa'+\gamma\gamma'\}}{\kappa'^2+\gamma'^2}$$

$$\times\left(\underline{\boldsymbol{w}}_{m's'\iota}(\kappa',\gamma',t)\underline{\boldsymbol{n}}^*_{m'\iota}(\kappa',\gamma',t)\right)$$

$$+\frac{q_{\iota}ek_0^2}{4\sqrt{\gamma_0}}\delta_{m,0}\delta(\gamma-\gamma_0)\tfrac{1}{2}\{\delta(\kappa-\kappa_0)+\delta(\kappa+\kappa_0)\}\sum_{s'}\int'\sqrt{\gamma'}\frac{(ss'kk'+\kappa\kappa'+\gamma\gamma')}{\kappa'^2+\gamma'^2}$$

$$\times\sum_{m'=-(\gamma\lambda_D)^{-1}}^{(\gamma\lambda_D)^{-1}}\underline{\boldsymbol{w}}_{m's'\iota}(\kappa',\gamma',t)\underline{\boldsymbol{n}}^*_{m'\iota}(\kappa',\gamma',t)$$

A-24

$$\left[\!\left[\mathbb{P}_{\sim\sim}^{(2)}\right]\!\right]_{\iota} = \frac{q_{\iota}ek^{2}}{4\sqrt{2}\sqrt{\gamma(\gamma+\gamma_{0})}}i\delta_{m,0}\tfrac{1}{2}\{\delta(\kappa-\kappa_{0})+\delta(\kappa+\kappa_{0})\}\sum_{m'=-((\gamma+\gamma_{0})\lambda_{d}/2)^{-1}}^{((\gamma+\gamma_{0})\lambda_{d}/2)^{-1}}(-1)^{m'}$$

$$\int'\delta(\gamma'-(\gamma+\gamma_{0})/2)\frac{(\kappa\gamma'-\kappa'\gamma)}{\sqrt{\kappa'^{2}+\gamma'^{2}}}\left(\underset{\sim}{\mathbf{u}}_{m'\iota}(\kappa',\gamma',t)\underset{\sim}{\mathbf{n}}_{m'\iota}^{*}(\kappa',\gamma',t)\right)$$

$$+\frac{q_{\iota}ek_{0}^{2}}{4\sqrt{\gamma_{0}}}i\delta_{m,0}\tfrac{1}{2}\{\delta(\kappa-\kappa_{0})+\delta(\kappa+\kappa_{0})\}\delta(\gamma-\gamma_{0})\int'\frac{(\kappa\gamma'-\kappa'\gamma_{0})\sqrt{\gamma'^{-1}}}{\sqrt{\kappa'^{2}+\gamma'^{2}}}$$

$$\times\sum_{m'=-(\gamma'\lambda_{D})^{-1}}^{(\gamma'\lambda_{D})^{-1}}\underset{\sim}{\mathbf{u}}_{m'\iota}(\kappa',\gamma',t)\underset{\sim}{\mathbf{n}}_{m'\iota}^{*}(\kappa',\gamma',t)$$

A-25

Equation 21: $\underset{\sim}{\mathbf{J}}_{m}^{I}(\sigma,t) = [\![\_\_]\!] + \sum_{\iota}\left(\left[\!\left[\mathbb{P}_{\sim\sim}^{(3)}\right]\!\right]_{\iota} + \left[\!\left[\mathbb{P}_{\sim\sim}^{(4)}\right]\!\right]_{\iota}\right)$

$$\left[\!\left[\mathbb{P}_{\sim\sim}^{(3)}\right]\!\right]_{\iota} = \frac{q_{\iota}e}{2\sqrt{2}}\sqrt{\frac{\gamma}{\gamma+\gamma_{0}}}k\delta_{m,0}\tfrac{1}{2}\{\delta(\kappa-\kappa_{0})+\delta(\kappa+\kappa_{0})\}\sum_{m'=-((\gamma+\gamma_{0})\lambda_{D}/2)^{-1}}^{((\gamma+\gamma_{0})\lambda_{D}/2)^{-1}}(-1)^{m'}$$

$$\int'\delta(\gamma'-(\gamma+\gamma_{0})/2)\frac{(\gamma\gamma'+\kappa\kappa')}{\sqrt{\kappa'^{2}+\gamma'^{2}}}\left(\underset{\sim}{\mathbf{u}}_{m'\iota}(\kappa',\gamma',t)\underset{\sim}{\mathbf{n}}_{m'\iota}^{*}(\kappa',\gamma',t)\right)$$

$$+\frac{q_{\iota}e}{2}\sqrt{\gamma_{0}}k_{0}\delta_{m,0}\tfrac{1}{2}\{\delta(\kappa-\kappa_{0})+\delta(\kappa+\kappa_{0})\}\delta(\gamma-\gamma_{0})\int'\sqrt{\gamma'^{-1}}\sum_{m'=-(\gamma'\lambda_{D})^{-1}}^{(\gamma'\lambda_{D})^{-1}}\frac{(\gamma\gamma'+\kappa\kappa')}{\sqrt{\kappa'^{2}+\gamma'^{2}}}$$

$$\times\left(\underset{\sim}{\mathbf{u}}_{m'\iota}(\kappa',\gamma',t)\underset{\sim}{\mathbf{n}}_{m'\iota}^{*}(\kappa',\gamma',t)\right)$$

A-26

$$\left[\!\!\left[\mathbb{P}^{(4)}_{\underset{\sim\sim}{}}\right]\!\!\right]_{\iota} = +i\frac{q_{\iota}e}{4\sqrt{2}}\sqrt{\gamma(\gamma+\gamma_0)}k\delta_{m,0}\tfrac{1}{2}\{\delta(\kappa-\kappa_0)+\delta(\kappa+\kappa_0)\}\sum_{m'=-((\gamma+\gamma_0)\lambda_D/2)^{-1}}^{((\gamma+\gamma_0)\lambda_D/2)^{-1}}(-1)^{m'}$$

$$\int'\delta(\gamma'-(\gamma+\gamma_0)/2)\frac{(\kappa\gamma'-\kappa'\gamma)}{\kappa'^2+\gamma'^2}\sum_{s'}\left(\underset{\sim}{\boldsymbol{w}}_{m's'\iota}(\kappa',\gamma',t)\underset{\sim}{\boldsymbol{n}}^{*}_{m'\iota}(\kappa',\gamma',t)\right)$$

$$+\frac{q_{\iota}e\sqrt{\gamma_0}}{2}k_0 i\delta_{m,0}\tfrac{1}{2}\{\delta(\kappa-\kappa_0)+\delta(\kappa+\kappa_0)\}\delta(\gamma-\gamma_0)\int'\frac{(\kappa\gamma'-\kappa'\gamma_0)\sqrt{\gamma'}}{\kappa'^2+\gamma'^2}$$

$$\times\sum_{m'=-(\gamma\lambda_D)^{-1}}^{(\gamma\lambda_D)^{-1}}\sum_{s'}\underset{\sim}{\boldsymbol{w}}_{m's'\iota}(\kappa',\gamma',t)\underset{\sim}{\boldsymbol{n}}^{*}_{m'\iota}(\kappa',\gamma',t)$$

A-27

Separation of Equation 22 is given in the main text.

Equation 23

$$\frac{\partial\underset{\sim}{\boldsymbol{w}}_{ms\iota}(\sigma,t)}{\partial t}-\frac{q_{\iota}e}{m_{\iota}}\underset{\sim}{\boldsymbol{E}}_{ms}(\sigma,t)+\frac{\eta q_{\iota}e}{m_{\iota}}\underset{\sim}{\boldsymbol{J}}^{S}_{ms}(\sigma,t)+\underset{\sim}{\boldsymbol{p}}^{S}_{ms\iota}(\sigma,t)$$

$$=\left[\!\!\left[--\right]\!\!\right]+\left[\!\!\left[\mathbb{P}^{(8)}_{\underset{\sim\sim}{}}\right]\!\!\right]_{\iota}+\left[\!\!\left[\mathbb{P}^{(9)}_{\underset{\sim\sim}{}}\right]\!\!\right]_{\iota}+\left[\!\!\left[\mathbb{P}^{(10)}_{\underset{\sim\sim}{}}\right]\!\!\right]_{\iota}+\left[\!\!\left[\mathbb{P}^{(11)}_{\underset{\sim\sim}{}}\right]\!\!\right]_{\iota}+\left[\!\!\left[\mathbb{P}^{(12)}_{\underset{\sim\sim}{}}\right]\!\!\right]_{\iota}+\left[\!\!\left[\mathbb{P}^{(13)}_{\underset{\sim\sim}{}}\right]\!\!\right]_{\iota}$$

A-28

$$\left[\!\!\left[\mathbb{P}^{(8)}_{\underset{\sim\sim}{}}\right]\!\!\right]_{\iota}=-\frac{ik^2\delta_{m,0}}{16\sqrt{\gamma}}\{\delta(\kappa-\kappa_0)+\delta(\kappa+\kappa_0)\}\int'\delta(\gamma'-(\gamma+\gamma_0)/2)k'^{-4}\gamma'^{2.5}$$

$$\times\sum_{m'=-(\gamma\lambda_D)^{-1}}^{(\gamma\lambda_D)^{-1}}\sum_{s'}(-1)^{m'}\underset{\sim}{\boldsymbol{w}}_{m's'\iota}(\kappa',\gamma',t)\underset{\sim}{\boldsymbol{w}}^{*}_{m's'\iota}(\kappa',\gamma',t)\left\{2\kappa'(ss'kk'-\kappa'\kappa+\gamma\gamma')-\frac{1}{2}k'^2\kappa\right\}$$

$$+\frac{ik^2\kappa_0}{16}\sqrt{\gamma_0}\delta_{m,0}\{\delta(\kappa-\kappa_0)-\delta(\kappa+\kappa_0)\}\delta(\gamma-\gamma_0)\int'k'^{-2}\gamma'^{1.5}$$

$$\times\sum_{m'=-(\gamma\lambda_D)^{-1}}^{(\gamma\lambda_D)^{-1}}\sum_{s'}\underset{\sim}{\boldsymbol{w}}_{m's'\iota}(\kappa',\gamma',t)\underset{\sim}{\boldsymbol{w}}^{*}_{m's'\iota}(\kappa',\gamma',t)$$

A-29

$$\left[\!\!\left[\mathbb{P}^{(9)}_{\underset{\sim\sim}{}}\right]\!\!\right]_{\iota}=\frac{k^2}{4\sqrt{\gamma}}\delta_{m,0}\tfrac{1}{2}\{\delta(\kappa-\kappa_0)+\delta(\kappa-\kappa_0)\}\int'\kappa'\gamma'^{1.5}k'^{-3}(\kappa'\gamma+\gamma'\kappa)\delta(\gamma'-(\gamma+\gamma_0)/2)$$

$$\times\sum_{m'=-(\gamma\lambda_D)^{-1}}^{(\gamma\lambda_D)^{-1}}\sum_{s'}\underset{\sim}{\boldsymbol{w}}_{m's'\iota}(\kappa',\gamma',t)\underset{\sim}{\boldsymbol{u}}^{*}_{m'\iota}(\kappa',\gamma',t)$$

A-30

$$\left[\!\left[\mathbb{P}_{\sim\sim}^{(10)}\right]\!\right]_\iota = +\frac{k^2}{16\sqrt{\gamma}}\delta_{m,0}\{\delta(\kappa-\kappa_0)+\delta(\kappa+\kappa_0)\}\int'\sqrt{\gamma'}k'^{-3}\delta(\gamma'-(\gamma+\gamma_0)/2)$$

$$\times\sum_{s'}\left((\gamma'^2-\kappa'^2)(ss'kk'-\kappa'\kappa)-\kappa'^2\gamma\gamma'\right)\sum_{m'=-(\gamma'\lambda_D)^{-1}}^{(\gamma'\lambda_D)^{-1}}(-1)^{m'}\underset{\sim}{\mathbf{u}}_{m'\iota}(\kappa',\gamma',t)\underset{\sim}{\mathbf{w}}_{m's'\iota}^*(\kappa',\gamma',t)$$

$$+\frac{k^2}{8\sqrt{\gamma}}\delta_{m,0}\{\delta(\kappa-\kappa_0)+\delta(\kappa+\kappa_0)\}\delta(\gamma-\gamma_0)\int'k'^{-3}\sqrt{\gamma'}$$

$$\sum_{s'}\left((\gamma'^2+\kappa'^2)(ss'kk'-\kappa'\kappa)-\kappa'^2\gamma\gamma'\right)\sum_{m'=-(\gamma'\lambda_D)^{-1}}^{(\gamma'\lambda_D)^{-1}}\underset{\sim}{\mathbf{u}}_{m'\iota}(\kappa',\gamma',t)\underset{\sim}{\mathbf{w}}_{m's'\iota}^*(\kappa',\gamma',t)$$

A-31

$$\left[\!\left[\mathbb{P}_{\sim\sim}^{(11)}\right]\!\right]_\iota = +\frac{k^2}{16\sqrt{\gamma}}i\delta_{m,0}\{\delta(\kappa-\kappa_0)+\delta(\kappa+\kappa_0)\}\int'\delta(\gamma'-(\gamma+\gamma_0)/2)$$

$$\times\sqrt{\gamma'^{-1}}(1-2k'^{-2}\kappa'^2)(\kappa'\gamma+\kappa\gamma')\sum_{m'=-\infty}^{\infty}(-1)^{m'}\underset{\sim}{\mathbf{u}}_{m'\iota}(\kappa',\gamma',t)\underset{\sim}{\mathbf{u}}_{m'\iota}^*(\kappa',\gamma',t)$$

$$+\frac{k^2}{4\sqrt{\gamma}}i\delta_{m,0}\tfrac{1}{2}\{\delta(\kappa-\kappa_0)+\delta(\kappa+\kappa_0)\}\delta(\gamma-\gamma_0)$$

$$\times\int'(\gamma'\kappa-\kappa'\gamma)\sqrt{\gamma'^{-1}}\sum_{m'=-\infty}^{\infty}\underset{\sim}{\mathbf{u}}_{m'\iota}(\kappa',\gamma',t)\underset{\sim}{\mathbf{u}}_{m'\iota}^*(\kappa',\gamma',t)$$

A-32

$$\left[\!\left[\mathbb{P}_{\sim\sim}^{(12)}\right]\!\right]_\iota = -\frac{q_\iota e}{m_\iota}i\frac{k^2}{4\sqrt{\gamma}}\delta_{m,0}\tfrac{1}{2}\{\delta(\kappa-\kappa_0)+\delta(\kappa+\kappa_0)\}\int'\delta(\gamma'-(\gamma+\gamma_0)/2)k'^{-4}\kappa'\gamma'^{1.5}$$

$$\times\sum_{m'=-(\gamma'\lambda_D)^{-1}}^{(\gamma'\lambda_D)^{-1}}(-1)^{m'}\sum_{s'}(sk\gamma'+s'k'\gamma)\underset{\sim}{\mathbf{w}}_{m's'\iota}(\kappa',\gamma',t)\underset{\sim}{\mathbf{B}}_{m's'}^*(\kappa',\gamma',t)$$

$$-i\frac{q_\iota e}{m_\iota}\frac{k_0^2}{4\sqrt{\gamma}}\delta_{m,0}\delta(\gamma-\gamma_0)\{\delta(\kappa-\kappa_0)+\delta(\kappa+\kappa_0)\}\int'k'^{-3}\gamma'^{1.5}(\kappa\gamma'-\kappa'\gamma)$$

$$\times\sum_{m'=-(\gamma'\lambda_D)^{-1}}^{(\gamma'\lambda_D)^{-1}}\sum_{s'}s'\underset{\sim}{\mathbf{w}}_{m's'\iota}(\kappa',\gamma',t)\underset{\sim}{\mathbf{B}}_{m's'}^*(\kappa',\gamma',t)$$

A-33

$$\left[\!\left[\mathbb{P}_{\sim\sim}^{(13)}\right]\!\right]_\iota = +\frac{q_\iota e}{m_\iota}\frac{k^2}{16\sqrt{\gamma}}\delta_{m,0}\{\delta(\kappa-\kappa_0)+\delta(\kappa+\kappa_0)\}\int'\delta(\gamma'-(\gamma+\gamma_0)/2)\sqrt{\gamma'}$$

$$\times \sum_{m'=-(\gamma'\lambda_D)^{-1}}^{(\gamma'\lambda_D)^{-1}} (-1)^{m'} \sum_{s'} \underline{u}_{m'\iota}(\kappa',\gamma',t)\, \underline{B}^*_{m's'}(\kappa',\gamma',t)\left[skk'^{-1}\left(1-2k'^{-2}\kappa'^2\right)+s'k'^{-2}(\gamma\gamma'+\kappa\kappa')\right]$$

$$+\frac{q_\iota e}{m_\iota}\frac{k^2}{4\sqrt{\gamma}}\delta_{m,0}\tfrac{1}{2}\{\delta(\kappa-\kappa_0)+\delta(\kappa+\kappa_0)\}\delta(\gamma-\gamma_0)$$

$$\times \int'\sqrt{\gamma'}\sum_{s'}\sum_{m'=-(\gamma'\lambda_D)^{-1}}^{(\gamma'\lambda_D)^{-1}} \underline{u}_{m'\iota}(\kappa',\gamma',t)\, \underline{B}^*_{m's'}(\kappa',\gamma',t)\left[skk'^{-1}-s'k'^{-2}(\gamma\gamma'+\kappa\kappa')\right]$$

A-34

Equation 24

$$\frac{\partial \underline{u}_{m\iota}(\sigma,t)}{\partial t}+\frac{kq_\iota e}{m_\iota}\underline{f}_m(\sigma,t)+\frac{\eta q_\iota e}{m_\iota}\underline{J}^I_m(\sigma,t)+\underline{p}^I_{m\iota}(\sigma,t)$$

$$=[\![\,-\,-\,]\!]+\left[\!\left[\mathbb{P}_{\sim\sim}^{(14)}\right]\!\right]_\iota+\left[\!\left[\mathbb{P}_{\sim\sim}^{(15)}\right]\!\right]_\iota+\left[\!\left[\mathbb{P}_{\sim\sim}^{(16)}\right]\!\right]_\iota+\left[\!\left[\mathbb{P}_{\sim\sim}^{(17)}\right]\!\right]_\iota+\left[\!\left[\mathbb{P}_{\sim\sim}^{(18)}\right]\!\right]_\iota+\left[\!\left[\mathbb{P}_{\sim\sim}^{(19)}\right]\!\right]_\iota$$

A-35

Using polytropic EoS with index 2: $\underline{p}^I_{m\iota}(\sigma,t)=2k\underline{n}(\sigma,t)\dfrac{p_0}{m_\iota n_0^2}$

$$\left[\!\left[\mathbb{P}_{\sim\sim}^{(14)}\right]\!\right]_\iota = +\frac{\sqrt{\gamma}}{8}\delta_{m,0}k\{\delta(\kappa-\kappa_0)+\delta(\kappa+\kappa_0)\}\int'\delta(\gamma'-(\gamma+\gamma_0)/2)\gamma'^{1.5}k'^{-4}$$

$$\times\left[\kappa'\gamma(2\kappa'\gamma')+\kappa\gamma\gamma'(s'k'-\kappa')\right]\sum_{m'=-(\gamma'\lambda_D)^{-1}}^{(\gamma'\lambda_D)^{-1}}(-1)^{m'}\sum_{s'}\underline{w}_{m's'\iota}(\kappa',\gamma',t)\,\underline{w}^*_{m's'\iota}(\kappa',\gamma',t)$$

$$-\frac{\sqrt{\gamma_0}}{4}k_0\kappa_0\delta_{m,0}\delta(\gamma-\gamma_0)\{\delta(\kappa-\kappa_0)-\delta(\kappa+\kappa_0)\}\int'k'^{-4}\gamma'^{3.5}(s'k'+\kappa')$$

$$\times\sum_{s'}\sum_{m'=-(\gamma'\lambda_D)^{-1}}^{(\gamma'\lambda_D)^{-1}}\underline{w}_{m's'\iota}(\kappa',\gamma',t)\,\underline{w}^*_{m's'\iota}(\kappa',\gamma',t)$$

A-36

$$\left[\!\left[\mathbb{P}_{\sim\sim}^{(15)}\right]\!\right]_\iota = -\frac{i\delta_{m,0}\sqrt{\gamma}k}{4}\{\delta(\kappa-\kappa_0)+\delta(\kappa+\kappa_0)\}$$

$$\times \int'(\gamma\gamma'-\kappa'\kappa)\kappa'\gamma'^{1.5}k'^{-3}\delta(\gamma'-(\gamma+\gamma_0)/2)\sum_{m'=-(\gamma'\lambda_D)^{-1}}^{(\gamma'\lambda_D)^{-1}}(-1)^{m'}\sum_{s'}\underset{\sim}{\mathbf{w}}_{m's'\iota}(\kappa',\gamma',t)\underset{\sim}{\mathbf{u}}^*_{m'\iota}(\kappa',\gamma',t)$$

A-37

$$\left[\!\left[\mathbb{P}_{\sim\sim}^{(16)}\right]\!\right]_\iota = -i\frac{\sqrt{\gamma}}{16k}\delta_{m,0}\{\delta(\kappa-\kappa_0)+\delta(\kappa+\kappa_0)\}\int'\delta(\gamma'-(\gamma+\gamma_0)/2)$$

$$\times\sqrt{\gamma'}k'\left(1-2k'^{-2}\kappa'\kappa'\right)(-2\gamma\kappa'+\kappa\gamma')\sum_{s'}\sum_{m'=-(\gamma'\lambda_D)^{-1}}^{(\gamma'\lambda_D)^{-1}}(-1)^{m'}\underset{\sim}{\mathbf{u}}_{m'\iota}(\kappa',\gamma',t)\underset{\sim}{\mathbf{w}}^*_{m's'\iota}(\kappa',\gamma',t)$$

$$+i\frac{\sqrt{\gamma}}{4k}\delta_{m,0}\tfrac{1}{2}\{\delta(\kappa-\kappa_0)+\delta(\kappa+\kappa_0)\}\delta(\gamma-\gamma_0)$$

A-38

$$\times\int'\sqrt{\gamma'}k'(2\gamma\kappa'+\kappa\gamma')\sum_{m'=-(\gamma'\lambda_D)^{-1}}^{(\gamma'\lambda_D)^{-1}}\sum_{s'}\underset{\sim}{\mathbf{u}}_{m'\iota}(\kappa',\gamma',t)\underset{\sim}{\mathbf{w}}^*_{m's'\iota}(\kappa',\gamma',t)$$

$$\left[\!\left[\mathbb{P}_{\sim\sim}^{(17)}\right]\!\right]_\iota = \delta_{m,0}\frac{1}{4\sqrt{2}}\{\delta(\kappa-\kappa_0)+\delta(\kappa+\kappa_0)\}\frac{\sqrt{\gamma}k}{\sqrt{(\gamma+\gamma_0)}}\int'\delta(\gamma'-(\gamma+\gamma_0)/2)$$

$$\times(\gamma\gamma'-\kappa'\kappa)\left(1-2k'^{-2}\kappa'^2\right)\sum_{m'=-(\gamma'\lambda_D)^{-1}}^{(\gamma'\lambda_D)^{-1}}(-1)^{m'}\underset{\sim}{\mathbf{u}}_{m'\iota}(\kappa',\gamma',t)\underset{\sim}{\mathbf{u}}^*_{m'\iota}(\kappa',\gamma',t)$$

$$+\delta_{m,0}\frac{1}{4}\sqrt{\gamma}k\{\delta(\kappa-\kappa_0)+\delta(\kappa+\kappa_0)\}\delta(\gamma-\gamma_0)$$

A-39

$$\times\int'\sqrt{\gamma'^{-1}}(\gamma\gamma'+\kappa'\kappa)\sum_{m'=-(\gamma'\lambda_D)^{-1}}^{(\gamma'\lambda_D)^{-1}}\underset{\sim}{\mathbf{u}}_{m'\iota}(\kappa',\gamma',t)\underset{\sim}{\mathbf{u}}^*_{m'\iota}(\kappa',\gamma',t)$$

$$\left[\!\left[\mathbb{P}^{(18)}_{\sim\sim}\right]\!\right]_\iota = +\frac{eq_\iota}{m_\iota}\frac{1}{8\sqrt{2}}(\gamma+\gamma_0)^{1.5}\kappa_0\sqrt{\gamma}k\delta_{m,0}\{\delta(\kappa-\kappa_0)-\delta(\kappa+\kappa_0)\}$$

$$\int'\delta(\gamma'-(\gamma+\gamma_0)/2)\kappa'k'^{-3}\sum_{s'}\sum_{m'=-(\gamma'\lambda_D)^{-1}}^{(\gamma'\lambda_D)^{-1}}(-1)^{m'}s'\underset{\sim}{w}_{m's'\iota}(\kappa',\gamma',t)\underset{\sim}{B}^*_{m's'}(\kappa',\gamma',t)$$

$$-\frac{eq_\iota}{m_\iota}\frac{\sqrt{\gamma}}{2}k\delta_{m,0}\{\delta(\kappa-\kappa_0)+\delta(\kappa+\kappa_0)\}\delta(\gamma-\gamma_0)\int'\gamma'^{1.5}k'^{-3}(\gamma\gamma'+\kappa'\kappa)$$

$$\times\sum_{s'}\sum_{m'=-(\gamma'\lambda_D)^{-1}}^{(\gamma'\lambda_D)^{-1}}s'\underset{\sim}{w}_{m's'\iota}(\kappa',\gamma',t)\underset{\sim}{B}^*_{m's'}(\kappa',\gamma',t)$$

A-40

$$\left[\!\left[\mathbb{P}^{(19)}_{\sim\sim}\right]\!\right]_\iota = \frac{eq_\iota}{m_\iota}\frac{i}{8\sqrt{2}}\sqrt{\gamma}\sqrt{(\gamma+\gamma_0)}k\delta_{m,0}\{\delta(\kappa-\kappa_0)+\delta(\kappa+\kappa_0)\}$$

$$\times\int'\delta(\gamma'-(\gamma+\gamma_0)/2)(\kappa\gamma'-\gamma\kappa')k'^{-2}\sum_{s'}\sum_{m'=-(\gamma'\lambda_D)^{-1}}^{(\gamma'\lambda_D)^{-1}}(-1)^{m'}\underset{\sim}{u}_{m'\iota}(\kappa',\gamma',t)s'\underset{\sim}{B}^*_{m's'}(\kappa',\gamma',t)$$

$$+\frac{eq_\iota}{m_\iota}\frac{i}{4}\sqrt{\gamma_0}k_0\delta_{m,0}\{\delta(\kappa-\kappa_0)+\delta(\kappa+\kappa_0)\}\delta(\gamma-\gamma_0)\int'k'^{-2}(\gamma\kappa'-\kappa\gamma')\sqrt{\gamma'}$$

$$\times\sum_{s'}\sum_{m'=-(\gamma'\lambda_D)^{-1}}^{(\gamma'\lambda_D)^{-1}}\underset{\sim}{u}_{m'\iota}(\kappa',\gamma',t)s'\underset{\sim}{B}^*_{m's'}(\kappa',\gamma',t)$$

A-41

Random parts:

Random parts of spectral time evolution equations form a system of linear first order equations, with coefficients dependent on non-random parts. Terms of these equations are labeled as $\left[\!\left[\mathbb{P}^{(n)}_{\sim\sim}\right]\!\right]$

Equations 16-19: not written as they are trivial.

Equation 20:

$$\underset{\sim}{\boldsymbol{J}}^{S}_{ms}(\sigma,t) = \sum_{\iota}\left(\left[\!\left[\mathbb{P}^{(1)}_{\sim\sim}\right]\!\right]_{\iota} + \left[\!\left[\mathbb{P}^{(2)}_{\sim\sim}\right]\!\right]_{\iota}\right) \qquad \text{A-42}$$

$$\left[\!\left[\mathbb{P}^{(1)}_{\sim\sim}\right]\!\right]_{\iota} = +\frac{q_{\iota}e}{2}\left(\sum_{s}\underset{\sim}{\boldsymbol{w}}_{ms\iota}(\kappa,\gamma,t)\right)\int''\underset{\sim}{\boldsymbol{n}}_{0\iota}(\kappa'',\gamma'',t)$$
$$\times\{\delta(\kappa''+\kappa_{0})+\delta(\kappa''-\kappa_{0})\}\{\delta(\gamma''-\gamma_{0})+\tfrac{1}{2}(-1)^{m}k^{-2}\gamma^{2}\delta(\gamma''-(2\gamma+\gamma_{0}))\}$$
$$+\frac{q_{\iota}e}{8\sqrt{\gamma}}\underset{\sim}{\boldsymbol{n}}_{m\iota}(\kappa,\gamma,t)\int'\{\delta(\kappa'+\kappa_{0})+\delta(\kappa'-\kappa_{0})\}\{\delta(\gamma'-\gamma_{0})+(-1)^{m}\delta(\gamma'-(2\gamma-\gamma_{0}))\}$$
$$\times\sqrt{\gamma'}k'^{-2}\sum_{s'}\underset{\sim}{\boldsymbol{w}}_{0s'\iota}(\kappa',\gamma',t)(ss'kk'+\kappa\kappa'+\gamma\gamma') \qquad \text{A-43}$$

$$\left[\!\left[\mathbb{P}^{(2)}_{\sim\sim}\right]\!\right]_{\iota} = +\frac{q_{\iota}ei}{8\gamma}\underset{\sim}{\boldsymbol{n}}_{m\iota}(\kappa,\gamma,t)\int'k'^{-1}(\underset{\sim}{\boldsymbol{u}}_{0\iota}(\kappa',\gamma',t))(\kappa\gamma'-\kappa'\gamma)\sqrt{\gamma\gamma'^{-1}}\{\delta(\kappa'+\kappa_{0})+\delta(\kappa'-\kappa_{0})\}$$
$$\times\{\delta(\gamma'-\gamma_{0})+(-1)^{m}\delta(\gamma'-(2\gamma-\gamma_{0}))\}$$
$$-(-1)^{m}\frac{q_{\iota}e}{4}i\kappa k^{-1}\underset{\sim}{\boldsymbol{u}}_{m\iota}(\kappa,\gamma,t)\int''\delta(\gamma''-(2\gamma+\gamma_{0}))\underset{\sim}{\boldsymbol{n}}_{0\iota}(\kappa'',\gamma'',t)\{\delta(\kappa''+\kappa_{0})+\delta(\kappa''-\kappa_{0})\} \qquad \text{A-44}$$

Equation 21:

$$\underset{\sim}{\boldsymbol{J}}^{I}_{m}(\sigma,t) = \sum_{\iota}\left(\left[\!\left[\mathbb{P}^{(3)}_{\sim\sim}\right]\!\right]_{\iota} + \left[\!\left[\mathbb{P}^{(4)}_{\sim\sim}\right]\!\right]_{\iota}\right) \qquad \text{A-45}$$

$$\left[\!\left[\mathbb{P}^{(3)}_{\sim\sim}\right]\!\right]_{\iota} = \frac{q_{\iota}e}{4}\underset{\sim}{\boldsymbol{u}}_{m\iota}(\kappa,\gamma,t)\sqrt{\gamma}k\int''\{\delta(\kappa''+\kappa_{0})+\delta(\kappa''-\kappa_{0})\}$$
$$\times\{2\delta(\gamma''-\gamma_{0})-(-1)^{m}(1-2k^{-2}\kappa^{2})\delta(\gamma''-(2\gamma+\gamma_{0}))\}\sqrt{\gamma'^{-1}}k'^{-1}\underset{\sim}{\boldsymbol{n}}_{0\iota}(\kappa'',\gamma'',t)$$
$$+\frac{q_{\iota}e}{4k}\underset{\sim}{\boldsymbol{n}}_{m\iota}(\kappa,\gamma,t)\sqrt{\gamma}\int'\{\delta(\kappa'+\kappa_{0})+\delta(\kappa'-\kappa_{0})\}\{\delta(\gamma'-\gamma_{0})+(-1)^{m}\delta(\gamma'-(2\gamma-\gamma_{0}))\}$$
$$\times\sqrt{\gamma'^{-1}}k'^{-1}(\gamma\gamma'+\kappa\kappa')\underset{\sim}{\boldsymbol{u}}_{0\iota}(\kappa',\gamma',t) \qquad \text{A-46}$$

$$\left[\!\left[\mathbb{P}_{\sim-}^{(4)}\right]\!\right]_\iota = +\frac{q_\iota e}{2}ik^{-3}\kappa\gamma^2(-1)^m\sum_{s'}\underset{\sim}{w}_{ms\iota}(\kappa,\gamma,t)$$

$$\times\int''\delta(\gamma''-(2\gamma+\gamma_0))\{\delta(\kappa''+\kappa_0)+\delta(\kappa''-\kappa_0)\}\underline{n}_{0\iota}(\kappa'',\gamma'',t)$$

$$+\frac{q_\iota e\sqrt{\gamma}}{4k}i\underset{\sim}{n}_{m\iota}(\kappa,\gamma,t)\int'\{\delta(\kappa'+\kappa_0)+\delta(\kappa'-\kappa_0)\}\{\delta(\gamma'-\gamma_0)+(-1)^m\delta(\gamma'-(2\gamma-\gamma_0))\}$$

$$\times k'^{-2}\sqrt{\gamma'}(\kappa\gamma'-\kappa'\gamma)\sum_{s'}\underset{\sim}{w}_{0s'\iota}(\kappa',\gamma',t)$$

A-47

Equation 22: given in main text.

Equation 23

$$\frac{\partial \underset{\sim}{w}_{ms\iota}(\sigma,t)}{\partial t} - \frac{q_\iota e}{m_\iota}\left(\underset{\sim}{E}_{ms}+\mathbb{F}(k)\exp(ickt)\right) + \frac{\eta q_\iota e}{m_\iota}\underset{\sim}{J}_{ms}^S(\sigma,t)$$

$$= \left[\!\left[\mathbb{P}_{\sim-}^{(8)}\right]\!\right]_\iota + \left[\!\left[\mathbb{P}_{\sim-}^{(9)}\right]\!\right]_\iota + \left[\!\left[\mathbb{P}_{\sim-}^{(10)}\right]\!\right]_\iota + \left[\!\left[\mathbb{P}_{\sim-}^{(11)}\right]\!\right]_\iota + \left[\!\left[\mathbb{P}_{\sim-}^{(12)}\right]\!\right]_\iota + \left[\!\left[\mathbb{P}_{\sim-}^{(13)}\right]\!\right]_\iota$$

A-48

$$\left[\!\left[\mathbb{P}_{\sim-}^{(8)}\right]\!\right]_\iota = -\underset{\sim}{w}_{ms\iota}(\kappa,\gamma,t)\frac{i}{8}k^{-2}\int'\sum_{s'}\underset{\sim}{w}_{0s'\iota}(\kappa',\gamma',t)k'^{-2}\{\delta(\kappa'+\kappa_0)+\delta(\kappa'-\kappa_0)\}$$

$$\times\left\{\begin{array}{l}\delta(\gamma'-\gamma_0)\left\{\begin{array}{l}\frac{1}{2}ss'kk_0\kappa\sqrt{\gamma_0}\left(3\gamma_0\sqrt{\gamma_0}-\gamma(\sqrt{\gamma_0}+\sqrt{\gamma})\right)\\+2\kappa'\gamma_0\gamma_0(k^2-2\gamma^2)+\sqrt{\gamma'\gamma}2k^2(\kappa'\gamma-\kappa\gamma')\end{array}\right\}\\+(-1)^m\delta(\gamma'-(2\gamma-\gamma_0))\left\{\begin{array}{l}(\kappa'\gamma+\kappa\gamma')\left(\sqrt{\gamma'\gamma}2k^2-\sqrt{\gamma'\gamma}2\gamma^2-ss'kk'\gamma'-\gamma'\kappa\kappa'-\gamma\gamma'\gamma'\right)\\+\frac{1}{2}ss'kk'\kappa\sqrt{\gamma'}(\gamma'\sqrt{\gamma'}-\gamma\sqrt{\gamma})\end{array}\right\}\end{array}\right\}$$

A-49

$$\left[\!\left[\mathbb{P}_{\sim-}^{(9)}\right]\!\right]_\iota = -\frac{k^{-2}}{8}\left(\sum_s\underset{\sim}{w}_{ms\iota}(\kappa,\gamma,t)\right)\int''k''^{-1}\underline{u}_{0v}(\kappa'',\gamma'',t)\{\delta(\kappa''+\kappa_0)+\delta(\kappa''-\kappa_0)\}$$

$$\{-2\delta(\gamma''-\gamma_0)(\gamma^2\kappa_0^2+\kappa^2\gamma_0^2)+(-1)^m\left(\kappa^2(2\gamma+\gamma_0)^2-\gamma^2\kappa_0^2\right)\delta(\gamma''-(2\gamma+\gamma_0))\}$$

$$+(-1)^m\frac{1}{4}\sqrt{\gamma}\kappa k^{-1}\underline{u}_{m\iota}(\kappa,\gamma,t)\int'\delta(\gamma'-(2\gamma-\gamma_0))\{\delta(\kappa'+\kappa_0)+\delta(\kappa'-\kappa_0)\}$$

$$\times k'^{-2}\sqrt{\gamma'}(\gamma\kappa'+\gamma'\kappa)\left(\sum_{s'}\underset{\sim}{w}_{0s'\iota}(\kappa',\gamma',t)\right)$$

A-50

$$\left[\!\left[\mathbb{P}^{(10)}_{\sim-}\right]\!\right]_\iota = \frac{1}{8}\gamma^{-1}k^{-1}\underline{\mathbf{u}}_{m\iota}(\kappa,\gamma,t)\sum_{s''}\int''\{\delta(\kappa''+\kappa_0)+\delta(\kappa''-\kappa_0)\}k''^{-2}\underline{\mathbf{w}}_{0s''\iota}(\kappa'',\gamma'',t)$$

$$\times\left\{2\gamma\gamma_0^2\left(ss''kk_0+2\kappa\kappa''\right)\delta(\gamma''-\gamma_0)-(-1)^m\left(ss''kk''(\gamma\gamma''-\kappa\kappa'')-\kappa^2\kappa_0^2\right)\gamma''\delta(\gamma''-(2\gamma+\gamma_0))\right\}$$

$$+\frac{\sqrt{\gamma}k^{-2}}{8}\underline{\mathbf{w}}_{ms\iota}(\kappa,\gamma,t)\int'\sqrt{\gamma'^{-1}}k'^{-1}\underline{\mathbf{u}}_{0\iota}(\kappa',\gamma',t)\{\delta(\kappa'+\kappa_0)+\delta(\kappa'-\kappa_0)\} \qquad \text{A-51}$$

$$\times\left\{\begin{array}{l}\left((\gamma'\gamma+\kappa'\kappa)(k^2+\kappa^2)+\kappa'\kappa\gamma^2\right)\delta(\gamma'-\gamma_0)\\ +(-1)^m\left((\gamma'\gamma-\kappa'\kappa)(k^2+\kappa^2)+\kappa'\kappa\gamma^2\right)\delta(\gamma'-(2\gamma-\gamma_0))\end{array}\right\}$$

$$\left[\!\left[\mathbb{P}^{(11)}_{\sim-}\right]\!\right]_\iota = -\frac{i}{8}\underline{\mathbf{u}}_{m\iota}(\kappa,\gamma,t)k^{-1}\int'\underline{\mathbf{u}}_{0\iota}(\kappa',\gamma',t)\{\delta(\kappa'+\kappa_0)+\delta(\kappa'-\kappa_0)\}$$

$$\times\left\{\begin{array}{l}2k'^{-1}\kappa\left(\kappa_0^2-\gamma_0^2\right)\delta(\gamma'-\gamma_0)\\ +(-1)^m k'^{-1}(\kappa'\kappa-\gamma\gamma')\left\{(\kappa'-\gamma^{-1}\gamma'\kappa)\delta(\gamma'-(2\gamma+\gamma_0))+2\kappa\sqrt{\gamma\gamma'^{-1}}\delta(\gamma'-(2\gamma-\gamma_0))\right\}\end{array}\right\} \qquad \text{A-52}$$

$$\left[\!\left[\mathbb{P}^{(12)}_{\sim-}\right]\!\right]_\iota = +\frac{q_\iota e}{m_\iota}\underline{\mathbf{w}}_{ms\iota}(\kappa,\gamma,t)\frac{i\kappa\gamma k^{-2}}{4}\sum_{s''}\int''\{\delta(\kappa''+\kappa_0)+\delta(\kappa''-\kappa_0)\}k''^{-2}\underline{\mathbf{B}}_{0s''}(\kappa'',\gamma'',t)$$

$$\times\left\{-2sk\gamma''\delta(\gamma''-\gamma_0)+(-1)^m(s''k''+sk)\gamma''\delta(\gamma''-(2\gamma+\gamma_0))\right\}$$

$$+\frac{q_\iota e}{m_\iota}\frac{i}{8}k^{-2}\left(\underline{\mathbf{B}}_{ms}(\kappa,\gamma,t)+isc^{-1}\mathbb{F}(k)\exp(ickt)\right) \qquad \text{A-53}$$

$$\times\sum_{s'}\int'k'^{-2}\sqrt{\gamma\gamma'}\underline{\mathbf{w}}_{0s'\iota}(\kappa',\gamma',t)\{\delta(\kappa'+\kappa_0)+\delta(\kappa'-\kappa_0)\}$$

$$\times\left\{\begin{array}{l}\left(2sk\kappa\gamma_0-sk\kappa'(\gamma_0+\gamma)+s'k'\kappa(\gamma-\gamma_0)\right)\delta(\gamma'-\gamma_0)\\ -\left(2sk\kappa\gamma'+sk\kappa'(\gamma-\gamma_0)+s'k'\kappa(3\gamma-\gamma_0)\right)(-1)^m\delta(\gamma'-(2\gamma-\gamma_0))\end{array}\right\}$$

$$\left[\!\left[\mathbb{P}^{(13)}_{\sim\sim}\right]\!\right]_\iota = +\frac{q_\iota e}{8m_\iota}\gamma^{-1}k^{-1}\underset{\sim}{u}_{m\iota}(\kappa,\gamma,t)\sum_{s''}\int''\{\delta(\kappa''+\kappa_0)+\delta(\kappa''-\kappa_0)\}k''^{-2}\underset{\sim}{B}_{0s''}(\kappa'',\gamma'',t)$$

$$\times\times\{2\gamma_0^2\gamma sk\delta(\gamma''-\gamma_0)-\gamma''[sk(\gamma\gamma''-\kappa\kappa'')+s''k''(\gamma\gamma-\kappa\kappa)](-1)^m\delta(\gamma''-(2\gamma+\gamma_0))\}$$

$$+\frac{q_\iota e}{m_\iota}\frac{1}{8}\sqrt{\gamma}k^{-1}s\left(\underset{\sim}{B}_{ms}(\kappa,\gamma,t)+isc^{-1}\mathbb{F}(k)\exp(ickt)\right) \qquad \text{A-54}$$

$$\times\int'\sqrt{\gamma'^{-1}}k'^{-1}\{\delta(\kappa'+\kappa_0)+\delta(\kappa'-\kappa_0)\}\underset{\sim}{u}_{0\iota}(\kappa',\gamma',t)$$

$$\times\{2(\gamma'\gamma+\kappa'\kappa)\delta(\gamma'-\gamma_0)-2\kappa'\kappa(-1)^m\delta(\gamma'-(2\gamma-\gamma_0))\}$$

Equation 24

$$\frac{\partial\underset{\sim}{u}_{m\iota}(\sigma,t)}{\partial t}+\frac{kq_\iota e}{m_\iota}\underset{\sim}{f}_m(\sigma,t)+\frac{\eta q_\iota e}{m_\iota}\underset{\sim}{J}^I_m(\sigma,t)+\underset{\sim}{p}^I_{m\iota}(\sigma,t)$$

$$=\left[\!\left[\mathbb{P}^{(14)}_{\sim\sim}\right]\!\right]_\iota+\left[\!\left[\mathbb{P}^{(15)}_{\sim\sim}\right]\!\right]_\iota+\left[\!\left[\mathbb{P}^{(16)}_{\sim\sim}\right]\!\right]_\iota+\left[\!\left[\mathbb{P}^{(17)}_{\sim\sim}\right]\!\right]_\iota+\left[\!\left[\mathbb{P}^{(18)}_{\sim\sim}\right]\!\right]_\iota+\left[\!\left[\mathbb{P}^{(19)}_{\sim\sim}\right]\!\right]_\iota \qquad \text{A-55}$$

$$\left[\!\left[\mathbb{P}^{(14)}_{\sim\sim}\right]\!\right]_\iota=\frac{1}{4}\sum_{s,s'}\underset{\sim}{w}_{ms\iota}(\kappa,\gamma,t)k^{-3}\int'\{\delta(\kappa'+\kappa_0)+\delta(\kappa'-\kappa_0)\}k'^{-2}\underset{\sim}{w}_{0s'\iota}(\kappa',\gamma',t)$$

$$\times\left\{\begin{array}{l}+\gamma^{2.5}\kappa\sqrt{\gamma'}\left(+s'k'\gamma-\kappa'\gamma+2\gamma_0\kappa\right)\delta(\gamma'-\gamma_0)\\ -(-1)^m\gamma\left(sk\kappa\gamma'^2-2\kappa\gamma\kappa'\gamma'-\gamma^2\kappa_0^2\right)\gamma'\delta(\gamma'-(2\gamma+\gamma_0))\\ -(-1)^m\gamma^{3.5}\kappa\sqrt{\gamma'}(s'k'+\kappa')\delta(\gamma'-(2\gamma-\gamma_0))\end{array}\right\} \qquad \text{A-56}$$

$$\left[\!\left[\mathbb{P}^{(15)}_{\sim\sim}\right]\!\right]_\iota=-\frac{1}{2}\gamma^2\kappa ik^{-3}\sum_s\underset{\sim}{w}_{ms\iota}(\kappa,\gamma,t)\int''\{\delta(\kappa''+\kappa_0)+\delta(\kappa''-\kappa_0)\}k''^{-1}\underset{\sim}{u}_{0\iota}(\kappa'',\gamma'',t)$$

$$\times\{\delta(\gamma''-\gamma_0)(\gamma''^2-\kappa''^2)-(\kappa\gamma''+\gamma\kappa'')(\gamma\gamma''+\kappa\kappa'')(-1)^m\delta(\gamma''-(2\gamma+\gamma_0))\}$$

$$-\frac{i}{4}\sqrt{\gamma}\underset{\sim}{u}_{m\iota}(\kappa,\gamma,t)\sum_{s'}\int'k'^{-2}\sqrt{\gamma'}\underset{\sim}{w}_{0s'\iota}(\kappa',\gamma',t)\{\delta(\kappa'+\kappa_0)+\delta(\kappa'-\kappa_0)\} \qquad \text{A-57}$$

$$\times\{(\kappa'\gamma-\kappa\gamma')\delta(\gamma'-\gamma_0)+(\kappa'\gamma+\kappa\gamma')(1-2k^{-2}\kappa^2)(-1)^m\delta(\gamma'-(2\gamma-\gamma_0))\}$$

$$\left[\!\left[\mathbb{P}_{\sim\sim}^{(16)}\right]\!\right]_{\iota} = -\frac{i}{8}k^{-2}\underset{\sim}{\mathbf{u}}_{m\iota}(\kappa,\gamma,t)\sum_{s''}\int''\gamma''k''^{-2}\left\{\delta(\kappa''+\kappa_0)+\delta(\kappa''-\kappa_0)\right\}\underset{\sim}{\mathbf{w}}_{0s''\iota}(\kappa'',\gamma'',t)$$

$$\times\left\{+2\gamma_0\kappa''\left(k^2+\gamma^2\right)\delta(\gamma''-\gamma_0)+(\kappa\kappa''-\gamma\gamma'')(2\gamma\kappa''-\kappa\gamma'')(-1)^{m'}\delta(\gamma''-(2\gamma+\gamma_0))\right\}$$

$$-\frac{i}{8}\gamma^{1.5}k^{-3}\sum_{s}\underset{\sim}{\mathbf{w}}_{0s\iota}(\kappa,\gamma,t)\int'\left\{\delta(\kappa'+\kappa_0)+\delta(\kappa'-\kappa_0)\right\}\sqrt{\gamma'^{-1}}k'^{-1}\underset{\sim}{\mathbf{u}}_{0\iota}(\kappa',\gamma',t)$$

$$\times\left\{(\gamma'\gamma+\kappa'\kappa)(2\gamma\kappa+\kappa\gamma)\delta(\gamma'-\gamma_0)+(\gamma'\gamma-\kappa'\kappa)(2\gamma\kappa-\kappa\gamma)(-1)^{m}\delta(\gamma'-(2\gamma-\gamma_0))\right\}$$

A-58

$$\left[\!\left[\mathbb{P}_{\sim\sim}^{(17)}\right]\!\right]_{\iota} = +\frac{1}{4}\underset{\sim}{\mathbf{u}}_{m\iota}(\kappa,\gamma,t)\int'k'^{-1}\left\{\delta(\kappa'+\kappa_0)+\delta(\kappa'-\kappa_0)\right\}\underset{\sim}{\mathbf{u}}_{0\iota}(\kappa',\gamma',t)$$

$$\times\left\{\begin{array}{l}\left(2\left(k^{-2}\gamma^2\gamma_0^2+k^{-2}\kappa^2\kappa'^2\right)+\sqrt{\gamma\gamma'^{-1}}(\gamma'\gamma+\kappa\kappa')\right)\delta(\gamma'-\gamma_0)\\ -(-1)^{m}\left(\gamma k^{-1}\gamma'+k^{-1}\kappa\kappa'\right)\left(\gamma k^{-1}\gamma'-\kappa'\kappa k^{-1}\right)\delta(\gamma'-(2\gamma+\gamma_0))\\ +(-1)^{m}\sqrt{\gamma\gamma'^{-1}}\left(1-2k^{-2}\kappa^2\right)(\gamma'\gamma-\kappa\kappa')\delta(\gamma'-(2\gamma-\gamma_0))\end{array}\right\}$$

A-59

$$\left[\!\left[\mathbb{P}_{\sim\sim}^{(18)}\right]\!\right]_{\iota} = +\frac{eq_{\iota}}{m_{\iota}}\frac{\gamma}{4}k^{-3}\sum_{s,s'}\underset{\sim}{\mathbf{w}}_{ms\iota}(\kappa,\gamma,t)\int'\left\{\delta(\kappa'+\kappa_0)+\delta(\kappa'-\kappa_0)\right\}k'^{-2}\underset{\sim}{\mathbf{B}}_{0s'}(\kappa',\gamma',t)$$

$$\times\left\{-2sk\gamma\gamma_0^2\delta(\gamma'-\gamma_0)+\left(sk(\gamma\gamma'+\kappa\kappa')+s'k'(\gamma\gamma-\kappa\kappa)\right)(-1)^{m}\gamma'\delta(\gamma'-(2\gamma+\gamma_0))\right\}$$

$$+\frac{eq_{\iota}}{m_{\iota}}\frac{1}{4}\gamma^{1.5}k^{-3}\sum_{s',s}\left(\underset{\sim}{\mathbf{B}}_{ms}(\kappa,\gamma,t)+isc^{-1}\mathbb{F}(k)\exp(ickt)\right)$$

$$\times\int'\sqrt{\gamma'}k'^{-2}\underset{\sim}{\mathbf{w}}_{0s'\iota}(\kappa',\gamma',t)\left\{\delta(\kappa'+\kappa_0)+\delta(\kappa'-\kappa_0)\right\}$$

$$\times\left\{\begin{array}{l}+\left\{sk(\gamma\gamma'+\kappa'\kappa)-s'k'(\gamma\gamma+\kappa\kappa)\right\}\delta(\gamma'-\gamma_0)\\ -\left\{s'k'(\gamma\gamma-\kappa\kappa)+sk(\kappa'\kappa+\gamma\gamma')\right\}(-1)^{m}\delta(\gamma'-(2\gamma-\gamma_0))\end{array}\right\}$$

A-60

$$\left[\!\left[\mathbb{P}^{(19)}_{\underset{\sim\sim}{}}\right]\!\right]_{\iota} = -(-1)^{m}\frac{eq_{\iota}}{m_{\iota}}\frac{1}{2}i\kappa\gamma k^{-2}\underset{\sim}{\mathbf{u}}_{m\iota}(\kappa,\gamma,t)\int'\gamma'k'^{-1}\left(\sum_{s'}s'\underline{\mathbf{\mathit{B}}}_{0s'}(\kappa',\gamma',t)\right)$$

$$\times\delta(\gamma'-(2\gamma+\gamma_{0}))\{\delta(\kappa'+\kappa_{0})+\delta(\kappa'-\kappa_{0})\}$$

$$+\frac{eq_{\iota}}{m_{\iota}}\frac{1}{4}i\gamma^{1.5}k^{-2}\left(\sum_{s}s\left(\underset{\sim}{\mathbf{\mathit{B}}}_{ms}(\kappa,\gamma,t)+isc^{-1}\mathbb{F}(k)\exp(ickt)\right)\right) \qquad \text{A-61}$$

$$\times\int'\sqrt{\gamma'^{-1}}k'^{-1}(\kappa\gamma'-\gamma\kappa')\underline{\mathbf{u}}_{0\iota}(\kappa',\gamma',t)$$

$$\times\{\delta(\kappa'+\kappa_{0})+\delta(\kappa'-\kappa_{0})\}\{\delta(\gamma'-\gamma_{0})-(-1)^{m}\delta(\gamma'-(2\gamma-\gamma_{0}))\}$$